\documentclass[iop]{emulateapj}

\usepackage{graphicx}
\usepackage{epstopdf}
\usepackage{epsfig}
\usepackage{mathtools}
\usepackage{natbib}
\usepackage{amsmath}
\usepackage{mathtools}

\slugcomment{}

\shorttitle{The Evolution of Gas in Transition Disks}
\shortauthors{Hoadley et al. 2015}

\begin{document}

\title{The Evolution of Inner Disk Gas in Transition Disks}
\accepted{\today}
\author{K. Hoadley\altaffilmark{1,2}, K. France\altaffilmark{1,2}, R. D. Alexander\altaffilmark{3}, M. McJunkin\altaffilmark{1,2}, and P. C. Schneider\altaffilmark{4}} 
\altaffiltext{1}{Laboratory for Atmospheric and Space Physics, University of Colorado, Boulder, CO 80303-7814}
\altaffiltext{2}{Center for Astrophysics and Space Astronomy, University of Colorado, Boulder, CO 80309-0389}
\altaffiltext{3}{Department of Physics \& Astronomy, University of Leicester, University Road, Leicester, LE1 7RH, UK}
\altaffiltext{4}{ESTEC/ESA, Keplerlaan 1, 2201 AZ Noordwijk, The Netherlands}
\email{keri.hoadley@colorado.edu}

\begin{abstract}
Investigating the molecular gas in the inner regions of protoplanetary disks provides insight into how the molecular disk environment  changes during the transition from primordial to debris disk systems. We conduct a small survey of molecular hydrogen (H$_2$) fluorescent emission, using 14 well-studied Classical T Tauri stars at two distinct dust disk evolutionary stages, to explore how the structure of the inner molecular disk changes as the optically thick warm dust dissipates. We simulate the observed HI-Lyman $\alpha$-pumped H$_2$ disk fluorescence by creating a 2D radiative transfer model that describes the radial distributions of H$_{2}$ emission in the disk atmosphere and compare these to observations from the Hubble Space Telescope. We find the radial distributions that best describe the observed H$_2$ FUV emission arising in primordial disk targets (full dust disk) are demonstrably different than those of transition disks (little-to-no warm dust observed). For each best-fit model, we estimate inner and outer disk emission boundaries (r$_{in}$ and r$_{out}$), describing where the bulk of the observed H$_2$ emission arises in each disk, and we examine correlations between these and several observational disk evolution indicators, such as n$_{13-31}$, r$_{in,CO}$, and the mass accretion rate. We find strong, positive correlations between the H$_2$ radial distributions and the slope of the dust SED, implying the behavior of the molecular disk atmosphere changes as the inner dust clears in evolving protoplanetary disks. Overall, we find that H$_2$ inner radii are $\sim$4 times larger in transition systems, while the bulk of the H$_2$ emission originates inside the dust gap radius for all transitional sources.

\end{abstract}

\keywords{stars: pre-main sequence - circumstellar matter - protoplanetary disks - molecules - line: profiles - ultraviolet: stars}


\section{Introduction}

Protoplanetary disks (PPDs) provide the raw materials for the formation of stellar systems \citep{Brown+09, Woitke+09b, Dullemond+Monnier+10}. Planet formation occurs near the midplane of a PPD, where column densities and optical depths are high \citep{Trilling+02, Armitage+03}, making it difficult to directly observe the material involved in the formation process \citep{Kominami+Ida+02}. Current understanding of the formation and evolution of planetary systems in gaseous disks comes from studies of molecular content above or near disk midplanes, which place limits on the composition and density distribution of the gas and dust content in the inner (r $\leq$ 10 AU) planet-forming regions \citep{Agundez+08, Carr+Najita+08, Carr+Najita+11, Salyk+08, Salyk+11b, Woitke+09b, Illacy+Woods+09, Heinzeller+11, Najita+11}. ``Transition'' disks refer to a class of PPDs with an optically thick outer zone but an inner region significantly depleted of dust grains \citep{Sato+99, Calvet+02, Salyk+09} and are traditionally identified by the deficiency in near-infrared (IR) flux and steep rise of far-IR flux in the observed SED \citep{Strom+89, Calvet+02, Calvet+05, Espaillat+07a}. Several theories exist for how dust gaps are opened in transition disks, including photoevaporation \citep{Hollenbach+94, Alexander+06, Alexander+13, Alexander+Armitage+07, Gorti+09}, dynamical clearing by protoplanetary systems \citep{Calvet+02, Rice+03, Dodson-Robinson+Salyk+11}, and dust grain growth \citep{Tanaka+05}. \\
\indent Discoveries of significant quantities of gas left within the dust gaps of transition disks (see \citealt{Najita+03, Rettig+04, Salyk+07}) and sharp ``walls'' between the thin and thick dust disk regions \citep{Brown+08} support the possibility of transition disks being carved out by giant planet formation and evolution \citep{Salyk+09, Dodson-Robinson+Salyk+11, Dong+14}. The remnant gas disks provide constraints on the processes that create the final structure of planetary systems, such as the transfer of gas from the PPD to circumplanetary disks, potentially leading to growth of protoplanets \citep{Lubow+99, Lubow+D'Angelo+06, Ayliffe+Bate+10, Beck+12}. Additionally, the molecular atmosphere of transition disks may respond to the dynamical perturbations caused by the presence of giant protoplanets and can lead to potentially observable effects, such as line asymmetries and distortions in near-IR CO emission profiles \citep{Regaly+10}. The strength of molecular emission originating from the inner radii of PPDs is dependent on the gas temperature, density, and degree of grain growth \citep{Salyk+11a}. Molecular line surveys therefore provide the opportunity for a broad examination of the gas distributions in circumstellar environments \citep{Brown+13}. \\
\indent Molecular hydrogen (H$_2$) has been measured to be $\sim 10^4$ times more abundant than any other molecule in the inner disks of PPDs \citep{France+14b}. Depending on the density, H$_2$ can survive at temperatures up to 5000 K \citep{Williams+00} and self-shields against UV radiation, making it robust to both collisional- and photo-dissociation \citep{Beckwith+78, Beckwith+82, Beckwith+83}. Molecular hydrogen provides a diagnostic for the spatial and structural extent of the warm molecular surface of PPDs \citep{Ardila+02, Herczeg+04, Yang+11}. While photo-excited H$_2$ does not interact strongly with evolving protoplanets, it traces the underlying distribution of gas at planet-forming radii \citep{Ardila+02, Herczeg+04, Herczeg+06, France+12a}. 
However, H$_2$ has proven difficult to observe in PPDs: cold H$_2$ (T $\sim$ 10 K) does not radiate efficiently because it has no permanent dipole \citep{Sternberg+89}, so IR ro-vibrational transitions are weak, making them difficult to observe from the ground. Therefore, studies of molecular material in disks typically rely on other tracers available in the near- and mid-IR, such as CO and H$_2$O, to estimate the molecular disk environment and mass of the underlying H$_2$ reservoir in disks. \\
\indent The strongest transitions of H$_2$ are found in the FUV (912 - 1700 \AA), where dipole-allowed electronic transitions are primarily photo-excited (``pumped'') by Ly$\alpha$ photons generated near the protostellar surface \citep{France+12b, Schindhelm+12b}. Warm H$_2$ (T $\gtrsim$ 1500K) has a significant population in excited vibration ($v$ = 1, 2) and rotation quantum states of the ground electronic band (X$^{1}\Sigma^{+}_{g}$) \citep{Shull+78}. When a Ly$\alpha$ photon interacts with a warm H$_2$ molecule in the correct ground-state population [$v$,$J$], the H$_2$ molecule absorbs the photon, exciting it to vibration levels ($v'$ $\rightarrow$ 0-4) of the first electronic band (B$^{1}\Sigma^{+}_{u}$). Since molecular hydrogen has strong (A$_{ul} \sim 10^{8} s^{-1}$; see \citealt{Abgrall+93}) electronic transitions in the FUV, the excited H$_2$ ``immediately'' decays back to the ground state, emitting a fluorescent photon, observed as an FUV emission line. The probability for an H$_2$ excitation-to-ground state transition to emit a photon with wavelength $\lambda$ depends on the branching ratio of the allowed transitions to the ground electronic state. The brightest H$_2$ emission lines arise from excited states [$v',J'$] = [1,4], [1,7], [0,1], and [0,2], which have absorption coincidences with Ly$\alpha$ within 0 and 600 km s$^{-1}$ of the Ly$\alpha$ line center, large oscillator strengths, and relatively low energy ground-state levels \citep{Herczeg+02, Herczeg+05}. The set of emission lines produced in the [$v',J'$] $\rightarrow$ [$v'',J''$] decay is refered to as a progression. \\
\indent Previous work on FUV fluorescent H$_2$ emission utilized basic profile fitting or small-sample parametric sets to estimate inner disk diagnostics, such as column density and temperature of the radiating molecular populations (see \citealt{Herczeg+04, France+12a, France+12b}). In this study, we create 2D radiative transfer models of PPD atmospheres to reproduce observed FUV H$_2$ emission lines. The models simulate a disk with radial temperature and  density distributions, which depend on physical parameters of the stellar system, such as the disk inclination angle and stellar Ly$\alpha$ radiation profile (taken from \citealt{Schindhelm+12b}). Using the four strongest H$_2$ progressions, we compare radiative transfer emission line models to the spectra of 14 CTTSs (8 primordial, 6 transition disks) observed with the \textit{Hubble Space Telescope} (\textit{HST})/Cosmic Origins Spectrograph (COS) and Space Telescope Imaging Spectrograph (STIS). The goal of this modeling work is to examine the relationship between the evolution of warm dust in PPDs and the radial distribution of H$_2$ in the disk atmosphere. We aim to understand how the spatial distribution of warm H$_2$ relate to the structure of the dust disk and other well-studied molecular disk tracers, such as carbon monoxide (CO) and water (H$_2$O). \\ 
%
%
\indent In \S 2, we present the targets, observations, and selection criteria of H$_2$ emission features explored in this work. In \S 3, we describe the forward modeling process for estimating the warm H$_2$ disk radiation fields, and in \S 4 we analyze how the best-fit models are determined and define metrics used to quantify the evolution of H$_2$ radiation for each PPD. In \S 5, we discuss how the modeled radiation distributions of fluorescing H$_2$ evolve in PPDs, comparing our results with observable warm dust disk evolution, mass accretion rates, and additional inner disk molecular tracers. Finally, we summarize how the gas disk structure correlates with the dissipation of warm dust grains as PPDs evolve to debris disks in \S 6. 

\section{Observations and H$_2$ Emission Line Selection}
\begin{deluxetable*}{l c c c c c c c l}\tablenum{1}
\tablecaption{Stellar Parameters
	\label{table:star_parms}}
\tablewidth{0pt}
\tablecolumns{9}
\tablehead{
	\colhead{Target} & \colhead{Spect.} & \colhead{M$_{\star}$} & \colhead{d} & 
	\colhead{A$_{v}$} & \colhead{i$_d$} & \colhead{Age} & \colhead{v$\sin$i} & 
	\colhead{ref.\tablenotemark{a}} \\
	                 & {Type}           & {(M$_{\sun}$)}        & {(pc)} &  
									  & {($^{\circ}$)}    &  {(Myr)}              & {(km s$^{-1}$)} & }
\startdata
	AA Tau & K7 & 0.8 & 140 & 0.5 & 75 & 6.4 $\pm$ 0.2 & 11.4 & 2,8,9,11,12,15,17,25 \\
	BP Tau & K7 & 0.73 & 140 & 0.5 & 30 & 5.9 $\pm$ 0.3 & 7.8 & 4,8,9,12,15,17,18,30 \\
	CS Cha & K6 & 1.05 & 160 & 0.8 & 60 & 6.4 $\pm$ 0.1 & ... & 5,6,15,19,22 \\
	DF Tau A & M2 & 0.19 & 140 & 0.6 & 85 & 6.3 $\pm$ 0.5 & 16.1 & 9,11,12,16,17,18 \\
	DM Tau & M1.5 & 0.5 & 140 & 0.0 & 35 & 6.6 $\pm$ 0.2 & 10.0 & 3,11,12,15,17,25 \\
	GM Aur & K5.5 & 1.20 & 140 & 0.1 & 55 & 6.9 $\pm$ 0.2 & 12.4 & 3,8,9,11,12,15,17,25 \\
	HN Tau A & K5 & 0.85 & 140 & 0.5 & 40 & 1.9 $\pm$ 0.9	& 52.8 & 7,9,11,18,23 \\
	LkCa15 & K3 & 0.85 & 140 & 0.6 & 49 & 6.4 $\pm$ 0.3 & 12.5 & 3,8,10,12,15,17,18 \\
	RECX 11 & K4 & 0.80 & 97 & 0.0 & 70 & 4.0 $\pm$ 1.5 & ... & 14,15,20,21 \\
	RECX 15 & M2 & 0.40 & 97 & 0.0 & 60 & 6.0 $\pm$ 1.0 & ... & 15,20,21,32 \\
	SU Aur	& G1 & 2.30 & 140 & 0.9	& 62 & 2.5 $\pm$ 0.9 &	65.0 & 1,4,9,11,18 \\
	TW Hya	& K6 & 0.60 & 54 & 0.0 & 4 & 10.0 $\pm$ 6.0 & 6.0 & 3,13,16,24,28,29,31 \\
	UX Tau A & K2 & 1.30 & 140 & 0.2 & 35 & 6.1 $\pm$ 0.3 & 25.4 & 3,8,11,17,19 \\
	V4046 Sgr & K5 & 1.75 & 83 & 0.0 & 34 & 6.9 $\pm$ 0.1 & 14.2(+13.7) & 25,27,28,29 
\enddata
\tablenotetext{a}{~(1) \citet{Akeson+02}; (2) \citet{Andrews+Williams+07}; (3) \citet{Andrews+11}; (4) \citet{Bouvier+90}; (5) \citet{Espaillat+07a}; (6) \citet{Espaillat+11}; (7) \citet{France+12b}; (8) \citet{Furlan+11}; (9) \citet{Gullbring+98}; (10) \citet{Hartmann+Soberblom+Stauffer+87}; (11) \citet{Hartmann+Stauffer+89}; (12) \citet{Hartmann+98}; (13) \citet{Herczeg+Hillenbrand+08}; (14) \citet{Ingleby+11}; (15) \citet{Ingleby+13}; (16) \citet{Johns-Krull+Valenti+01}; (17) \citet{Kenyon+94}; (18) \citet{Kraus+Hillenbrand+09}; (19) \citet{Lawson+96}; (20) \citet{Lawson+01}; (21) \citet{Lawson+04}; (22) \citet{Luhman+04}; (23) \citet{McJunkin+13}; (24) \citet{Pontoppidan+08}; (25) \citet{Quast+00}; (26) \citet{Ricci+10}; (27) \citet{Rodriguez+10}; (28) \citet{Rosenfeld+12}; (29) \citet{Rosenfeld+13}; (30) \citet{Simon+00}; (31) \citet{Webb+99}; (32) \citet{Woitke+13}. }
\end{deluxetable*}
\indent We sample a large collection of \textit{HST}-COS and \textit{HST}-STIS (for TW Hya) FUV H$_2$ data to understand the relative changes in the radiation distributions of H$_2$ arising from the inner regions of primordial and transition disks. The observations were obtained through the DAO of Tau guest observing program (PID 11616; PI - G. Herczeg), the COS Guaranteed Time Observing program (PIDs 11533 and 12036; PI - J. Green), and \textit{HST} Program GTO-8041 (PI - J. Linsky). The observations have been presented in previous literature (for examples, see \citealt{Herczeg+06, Ingleby+11, Yang+11, France+12b, France+14, Schindhelm+12a, Ardila+13}). \\
\begin{figure*}[htp]
	\center
	\includegraphics[angle=90, width=1.0\textwidth]{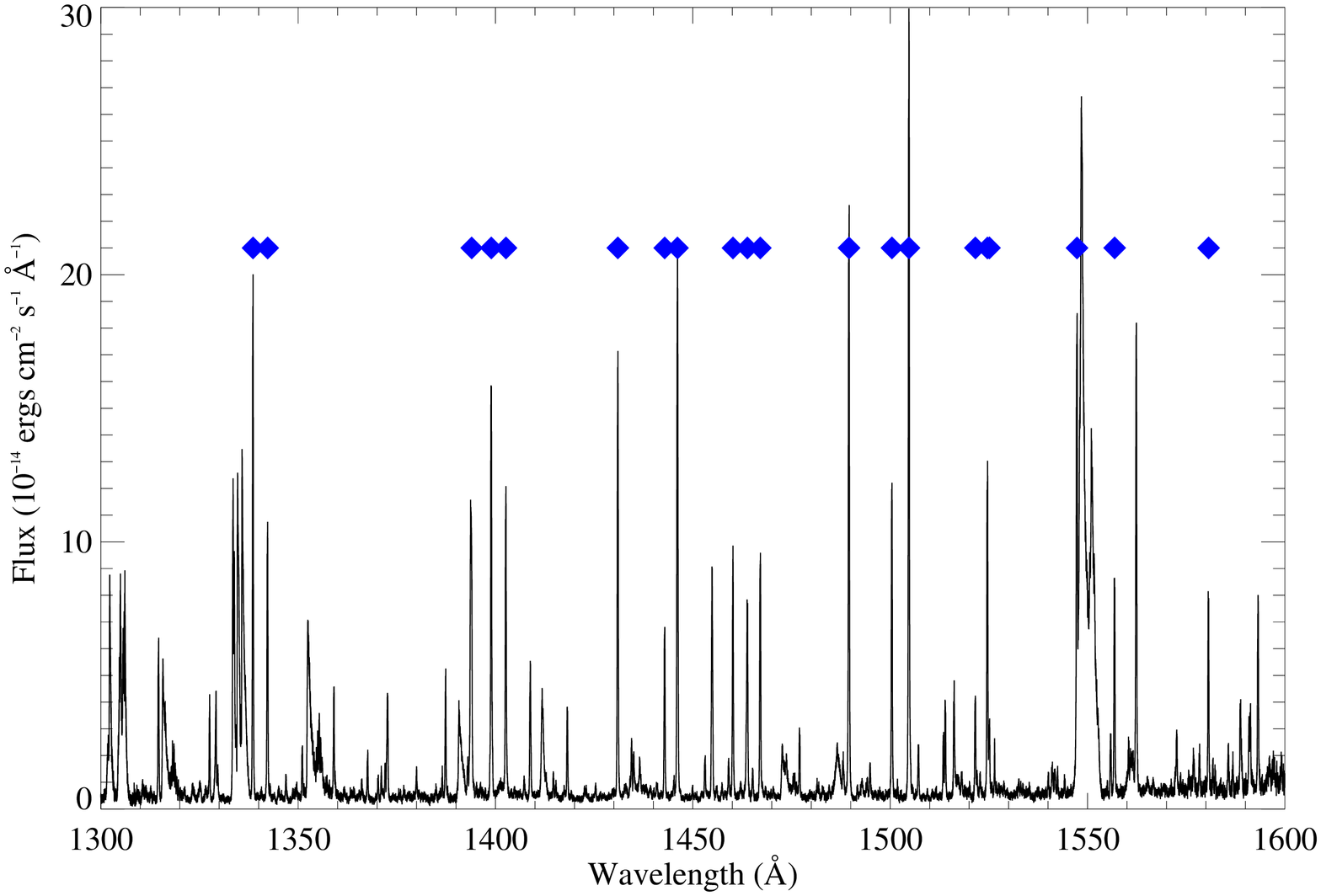} 
	\caption{The \textit{HST}-COS FUV spectrum of transition disk target V4046 Sgr, ranging from 1300 - 1600 {\AA}. Most of the narrow emission features are Ly$\alpha$ pumped fluorescent H$_2$ lines. Many of the strongest H$_2$ emission features fluoresce from Ly$\alpha$-pumped excited state progressions [1,4], [1,7], [0,1], or [0,2]. Blue diamonds mark the H$_2$ fluorescent emission lines studied in this work.}
	\label{spect_ex}
\end{figure*}
\indent The medium-resolution G130M and G160M FUV modes of COS \citep{Green+12} were utilized for all targets except TW Hya, which was observed with the E140M mode (1170 - 1710 \AA) with the 0.5\arcsec\ $\times$ 0.5\arcsec\ aperture of STIS at a resolving power of 25,000 (see \citealt{Herczeg+06}). The point-source resolution for each mode on COS is $\Delta v \approx$ 17 km s$^{-1}$ with 7 pixels per resolution element \citep{Osterman+11} and $\Delta v \approx$ 12 km s$^{-1}$ for the STIS E140M observing mode of TW Hya \citep{Leitherer+01}. The COS data were smoothed by 3 pixels for analysis. The one-dimensional spectra of COS were produced using the CALCOS COS calibration pipeline, which were aligned and coadded using a custom software procedure \citep{Danforth+10}. The STIS data were reduced using the CALSTIS STScI reduction pipeline \citep{Lindler+99}, with calibration lamp spectra obtained during observations to assign wavelength solutions. An example of the continuous far-UV spectrum of V4046 Sgr is shown in Figure~\ref{spect_ex}. \\
\indent  Stellar properties, such as mass, accretion rate, and inclination angle are used to constrain the underlying model framework. All disk inclination angles have been estimated from sub-mm/IR interferometric studies (see \citealt{Simon+00, Johns-Krull+Valenti+01, Andrews+Williams+07, Espaillat+07a, Andrews+11, Rosenfeld+12}). Stellar masses and extinction estimates were derived from pre-main sequence stellar evolutionary tracks \citep{Hartmann+98}. Mass accretion rates were estimated from measurements of the accretion luminosity \citep{Ingleby+13}. Refer to Table~\ref{table:star_parms} for lists of all the relevant stellar parameters, with references therein. \\ 
%
%
%
\begin{deluxetable*}{c c c c c c c c r}\tablenum{2}
\tabletypesize{\scriptsize}
\tablecaption{Selected H$_2$ Emission Lines \& Properties of H$_2$ Pumping Transitions
	\label{H2_prog}}
\tablecolumns{9}
\tablewidth{0pt}
\tablehead{
	\colhead{$\lambda_{lab}$} & \colhead{Progression} & \colhead{Line ID\tablenotemark{a}} & 
	\colhead{$\lambda_{pump}$}  &  \colhead{$v_{trans}$\tablenotemark{b}} & 
	\colhead{A$_{ul}$\tablenotemark{c}} & \colhead{$f$\tablenotemark{d}}  \\
	{(\AA)}                   &                       &                                    &   
  {(\AA)}                     & {(km s$^{-1}$)}                         & 
	{(10$^8$ s$^{-1}$)}                 & {(10$^{-3}$)} } 
\startdata
	1442.87 & [1,7] & $(1-6) R(6)$ & 1215.726 & 14 & 0.9 & 34.8 \\
	1467.08 &       & $(1-6) P(8)$ &          &    & 1.3 &      \\
	1500.45 &       & $(1-7) R(6)$ &          &    & 1.7 &      \\
	1524.65 &       & $(1-7) P(8)$ &          &    & 1.9 &      \\
	1556.87 &       & $(1-8) R(6)$ &          &    & 1.3 &      \\
	1580.67 &       & $(1-8) P(8)$ &          &    & 1.1 &      \\
	\hline 
	1431.01 & [1,4] & $(1-6) R(3)$ & 1216.070 & 99 & 1.0 & 28.9 \\
	1446.12 &       & $(1-6) P(5)$ &          &    & 1.4 &      \\
	1489.57 &       & $(1-7) R(3)$ &          &    & 1.6 &      \\
	1504.76 &       & $(1-7) P(5)$ &          &    & 2.0 &      \\
	1547.34 &       & $(1-8) R(3)$ &          &    & 1.1 &      \\
	\hline 
	1338.56 & [0,1] & $(0-4) P(2)$ & 1217.205 & 379 & 3.1 & 44.0 \\
	1398.95 &       & $(0-5) P(2)$ &          &    & 2.6 &      \\
	1460.17 &       & $(0-6) P(2)$ &          &    & 1.5 &      \\
	1521.59 &       & $(0-2) P(2)$ &          &    & 0.6 &      \\
	\hline 
	1342.26 & [0,2] & $(0-4) P(3)$ & 1217.643 & 487 & 2.8 & 28.9 \\
	1393.96 &       & $(0-5) R(1)$ &          &    & 1.6 &      \\
	1402.65 &       & $(0-5) P(3)$ &          &    & 2.3 &      \\
	1463.83 &       & $(0-6) P(3)$ &          &    & 1.4 &      \\
	1525.15 &       & $(0-7) P(3)$ &          &    & 0.5 &      
\enddata
\tablenotetext{a}{~Transitions are from the Lyman-excited to ground electronic states of the H$_2$ band system, $B^1 \Sigma^+_u - X^1 \Sigma^+_g$.}
\tablenotetext{b}{~Velocity from line center of the pumping transition of Ly$\alpha$.}
\tablenotetext{c}{~Einstein coefficient, describing the spontaneous decay rate from the electronically-excited Lyman band, taken from \citet{Abgrall+93}.}
\tablenotetext{d}{~Oscillator strengths from \citet{Abgrall+93}.}
\end{deluxetable*}
\indent The \textit{HST}-COS FUV spectra of all CTTSs from 1300 - 1600 {\AA} reveal a suite of H$_2$ fluorescence features linked to Ly$\alpha$-pumping. We chose to use the strongest transitions from the electronically-excited progressions [$v',J'$] = [0,1], [0,2], [1,4], and [1,7] for the purposes of studying the underlying general characteristics of the bulk gas disk. We sample 3 emission features from each progression. This gives us access to strong, non-blended emission lines that are well-defined from the FUV continuum, while balancing the CPU time required for detailed line profile analysis. We selected H$_2$ emission features by locating the strongest transitions for each progression, outlined by \citet{Herczeg+02}. 
See Table~\ref{H2_prog} for the full outline of transitions chosen.

\section{Modeling Analysis}

We create models of warm H$_2$ in PPDs to constrain the radial distribution of fluorescent H$_2$ emission in disk atmospheres. Our aim is to understand the relative changes in the H$_2$ distributions as we observe various stages of dust disk evolution. 
The fluorescent emission line shape and intensity depend on the physical conditions of the gas, while the observed line width depends predominantly on the disk inclination. We construct a physical model of the disk structure, motivated by the disk modeling analysis done by \citet{Rosenfeld+12}. \\
\indent The models make several basic assumptions on the disk properties: (a) the disk material orbits in Keplerian rotation around a central point mass, representing the stellar mass; (b) the H$_{2}$ fluorescence occurs in a shallow, warm layer on the disk surface; and (c) the level populations of warm H$_2$ that absorb the incident stellar Ly$\alpha$ radiation field are in local thermodynamic equilibrium (LTE). (a) implies that the gas disk mass is a small fraction of the stellar mass ($M_{d}/M_{\star} \ll 1$). Several studies have shown that the disk mass to stellar mass ratio ($M_d / M_{\star}$) $<$ 1\%, making this assumption plausible \citep{Andrews+13}. In the case of a binary system (i.e., V4046 Sgr), both stellar masses are represented as one central mass point. For (b), \citet{Herczeg+04} find that the warm H$_2$ disk layer interacting with the stellar Ly$\alpha$ to produced the observed fluorescence lines corresponds to mass column density of $\sim 10^{-5}$ g cm$^{-2}$, which is a much smaller mass column density  predicted to be within 1 AU by \citet{DAlessio+99}. This suggests that the Ly$\alpha$-pumped fluorescent emission originates from a tenuous layer of warm H$_2$ on the disk surface and supports a purely radial thermal distribution $T(r)$. 
For (c), combination of collisional excitation and radiative de-excitation is assumed to be in equilibrium to keep the H$_2$ gas near the disk surface at warm temperatures (T $>$ 1000 K; \citealt{Nomura+05, Nomura+07}). Previous studies of FUV H$_2$ emission have argued both for and against this assumption \citep{Ardila+02, Herczeg+06}. 
LTE conditions keep the assumed parameters straightforward and allow us to model the H$_2$ ground-state populations as a ``snapshot'' of the disk atmosphere as it was observed. \\
%
\indent The warm H$_2$ atmosphere is described by the surface density and temperature distribution of gas, which characterizes how much of the warm H$_2$ is populating excited ground-states [$v$,$J$]. We reference these physical quantities in cylindrical coordinate positions in the disk (\textit{r,$\phi$,z}). If we consider that a parcel of warm H$_{2}$ gas on the disk surface is characterized by its radial position, vertical height from the disk midplane, and velocity distribution (\textit{r, z}, $v_{\phi}(r)$), the velocity of the gas parcel, $v_{\phi}(r)$, is described by Keplerian rotation in $\hat{\phi}$ only:
\begin{equation}
	v_{\phi}(r) = v_{k} = \sqrt{\frac{G M_{\star}}{r}} ; v_{r} = v_{z} = 0 ,
	\label{v_phi}
\end{equation}
where $G$ is the gravitational constant and $M_{\star}$ is the central stellar mass. The mass density at the warm H$_2$ disk surface is a function of the radial and vertical height in the disk,
\begin{equation}
	\rho(r,z) = \frac{\Sigma (r)}{\sqrt{2 \pi} H_{p}} \exp{\left[-\frac{1}{2} \left(\frac{z}{H_{p}}\right)^{2}\right]} ,
	\label{rho_r_z}
\end{equation}
where $\Sigma (r)$ is the radial surface density distribution of H$_{2}$, and \textit{$H_{p}$} is the pressure scale height as a function of radius, defined as:
\begin{equation}
	H_{p} = \frac{c_{s}}{\Omega} = \sqrt{\frac{k T(r)}{\mu m_{H}}\cdot\frac{r^3}{G M_{\star}}} ,
	\label{H_p} 
\end{equation}
where $c_{s}$ is the sound speed, $\Omega$ is the angular velocity of the gas, $k$ is the Boltzmann constant, $T(r)$ is the radial temperature profile of the warm H$_{2}$ disk atmosphere, $\mu$ is the ``mean molecular weight'' of the gas, and $m_{H}$ is the mass of a hydrogen atom. The temperature distribution of the disk atmosphere is approximated as a power-law function:
\begin{equation}
	T(r) = T_{1 AU} \left(\frac{r}{1 AU} \right)^{-q} ,
	\label{T_r}
\end{equation}
where \textit{$T_{1 AU}$} is the temperature of the warm H$_2$ at r = 1 AU, and \textit{$q$} is the temperature gradient. \\
\indent We assume a radial surface density for a static accretion disk, represented by a power-law viscosity profile (see \citealt{Lynden-Bell+Pringle+74}),
\begin{equation}
	\Sigma(r) = \Sigma_{c} \left(\frac{r}{r_{c}}\right)^{- \gamma} \exp{\left[- \left(\frac{r}{r_{c}}\right)^{2 - \gamma}\right]} ,
	\label{Sigma_r}
\end{equation}
where \textit{$\gamma$} is the density gradient, \textit{$r_{c}$} is the characteristic radius of the gas in the disk, and $\Sigma_{c}$ is a normalization factor for the surface density distribution, dependent on the total H$_2$ mass contributing to the emission lines simulated by these models. The characteristic radius describes the transition from a power-law dominated density distribution to an exponentially-dominated density fall-off in the disk \citep{Lynden-Bell+Pringle+74, Hartmann+98}. It is important to note that $\Sigma(r)$ contains a normalization factor ($\Sigma_c$), which normalizes to the disk midplace density. Our models only attempt to describe the behavior of the disk atmosphere, where the warm, tenuous H$_2$ resides. As a consequence, the functionality of $\Sigma(r)$ serves as a structural layout of the radial H$_2$ disk atmosphere. Since we normalize $\Sigma(r)$ with a factor describing the disk midplane density, the solutions of $\Sigma(r)$ describe the radial distributions of warm H$_2$, but the resulting H$_2$ mass estimates are not meaningful. \\
\indent The level populations of warm, ground state H$_2$ contributing to the emission line are assumed to be in LTE and are determined using the Boltzmann equation,
\begin{equation}
\begin{split}
	n_{[v,J]}(r,z) = &\frac{\rho \left(r,z \right) X_{H_2}}{\mu m_{H}} \times  \\
	                & \frac{\text{g}_{[v,J]}}{Z_{[v,J]}\left(T \right)} \times \exp \left( \frac{- E_{[v,J]}}{kT \left(r \right)} \right) ,
	\label{n_l}
\end{split}
\end{equation}
where X$_{H_2}$ is the fraction of the total H$_2$ gas mass contributing to the fluorescence observed in the FUV, g$_{[v,J]}$ is the statistical weight of the level population, Z$_{[v,J]}$(T) is the partition function describing the likelihood that the warm H$_2$ is in state [$v$,$J$], and E$_{[v,J]}$ is the energy of warm H$_2$ in ground state [$v,J$]. \\
\indent The radial distribution of molecular hydrogen has two normalization factors ($X_{H_2}$ and $\Sigma_{c}$) that are not independent of disk conditions and are defined by their product in $n_{[v,J]}(r,z)$. The product of these factors describes the total mass of warm H$_2$ available for photo-excitation to state [$v'$,$J'$] ($M_{H_2}$), which is obtained by integrating the distribution over ($r$,$\phi$,$z$): \textit{$M_{H_2}$} = $X_{H_2} \Sigma_{c} \left(2 \pi r_{c}^2 \right) / \left(2 - \gamma \right)$. \\ 
%
%
\begin{figure}
	\includegraphics[angle=270, width=0.55\textwidth]{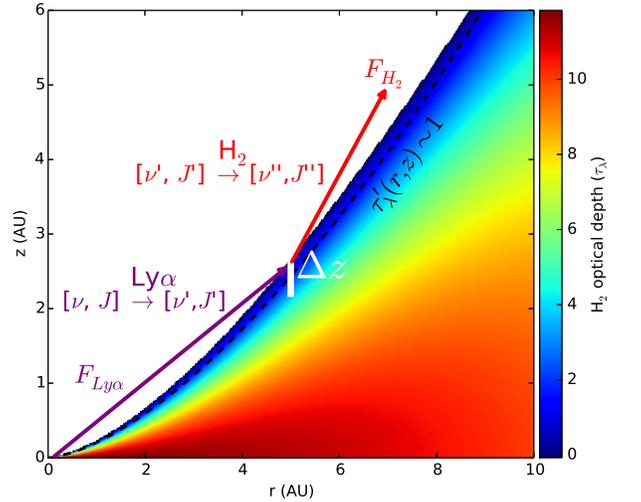}
	\caption{We provide a graphical representation of the H$_2$ disk atmosphere model. The disk contours represent the warm H$_2$ optical depth ($\tau_{\lambda}(r,z)$) to stellar Ly$\alpha$ radiation being pumped to state [$v'$,$J'$] = [1,4]. The dashed line marks off the approximate location of $\tau_{\lambda}^{'}$ $\approx$ 1, which is where the H$_2$ disk atmosphere becomes optically thick to the penetrating Ly$\alpha$ photons. The stellar Ly$\alpha$ radiation (purple arrow) is absorbed by the by the warm H$_2$, which is excited to state [$v'$,$J'$] and emits a photon ($\lambda_{H_2}$; red arrow) to decay back to ground state [$v''$,$J''$]. }
	\label{fig:model_rt}
\end{figure}
\indent The radiative transfer calculation required to reproduce the observed fluorescent H$_2$ emission happens in two steps: 1) the warm H$_2$ in ground state population [$X:v$,$J$] is pumped into a rovibrational level [$B:v'$,$J'$] of the excited electronic (Lyman band) state by the absorption of an incident stellar Ly$\alpha$ with wavelength $\lambda_{Ly\alpha}$, and 2) the excited H$_2$ molecule decays back to some ground electronic state [$X:v''$,$J''$], emitting a FUV photon with wavelength $\lambda_{H_2}$. Molecular hydrogen has an absorption cross section ($\sigma_{H_2}$) defined by the area around the molecule that can intersect an incoming photon with the appropriate energy for photo-excitation:
\begin{equation}
	\sigma_{H_2} = \frac{\lambda_{Ly\alpha}^3}{8 \pi c} \frac{\text{g}_{[B:v',J']}}{\text{g}_{[X:v,J]}} A_{lu} ,
	\label{x-sect}
\end{equation}
where $\lambda_{Ly\alpha}$ is the rest frame wavelength of the stellar Ly$\alpha$ line profile needed to excite the warm H$_2$ in ground state [$X:v,J$] up to energy level [$B:v',J'$], and $A_{lu}$ is the probability that H$_2$ in population [$X:v,J$] will be ``pumped'' to electronic state [$B:v',J'$]. Note that, for the remainder of this paper, we will omit the ground state H$_2$ ($X$) and excited state ($B$) level branch denominations from the vibration and rotation state discussion. \\
\begin{deluxetable}{c c c}\tablenum{3}
\tabletypesize{\scriptsize}
\tablecaption{Parametric Values Explored in Modeling Framework
	\label{tab:parm_values}}
\tablewidth{0pt}
\tablehead{
	\colhead{Parameter} & \colhead{Values} & \colhead{Units} }
\startdata
z/r	&	(2, 3, 4, 5, 6, 7) $\times$ H$_p$  	&  \\
\\
$\gamma$ 	&	0.0, 0.25, 0.5, 0.75, 1.25, 1.5, 1.75, 1.99  &  \\
\\
q					& -1.0, -0.5, -0.25, -0.1, -0.05,  &	 \\
					& 0.0, +0.05, +0.1, +0.25, +0.5    &   \\
\\
T$_{1AU}$	& 500, 1000, 1500, 2000, 2500,  & K \\
					& 3000, 3500, 4000, 4500, 5000  &   \\
\\
r$_{c}$   & 0.1, 0.5, 1.0, 3.0, 5.0, 7.5, 10.0, 20.0 & AU \\
\\
M$_{H_2}$ & 5$\times$10$^{-10}$, 10$^{-10}$, 5$\times$10$^{-11}$,  & M$_{\sun}$ \\ 
					&	10$^{-11}$, 5$\times$10$^{-12}$, 10$^{-12}$,  &    \\
					& 5$\times$10$^{-13}$, 10$^{-13}$, 5$\times$10$^{-14}$, 10$^{-14}$	&  
\enddata
\tablenotetext{}{Values were chosen to reproduce the desired H$_2$ features \citep{Herczeg+04, France+12b}. The only parameters without aforementioned constraints were \textit{z/r, $\gamma$, q} and \textit{r$_{c}$} because literature estimates of these values were not known. $\gamma$ and $q$ were constrained by the power-law functionality role they play in the models, and \textit{r$_c$} was estimated around $<$r$_{[1,7]}$$>$ calculated by \citet{France+12b}.}
\end{deluxetable}
%
%
%
\begin{deluxetable}{c c c c c}\tablenum{4}
\tabletypesize{\scriptsize}
\tablecaption{Minimum $\chi^2$ Statistics for Each Progression Fit
	\label{tab:min_chi2}}
\tablewidth{0pt}
\tablehead{
	 & & Progression &    [$v'$,$J'$]  &  \\
	 \colhead{Target} & \colhead{[0,1]} & \colhead{[0,2]} & \colhead{[1,4]} & \colhead{[1,7]} }
\startdata

	AA Tau (2011) & 5.37 & 6.48 & 5.52 & 1.25 \\
	AA Tau (2013) & 1.78 & 5.29 & 4.24 & 1.62 \\
	BP Tau & 2.82 & 51.75 & 5.28 & 2.97 \\
	CS Cha & 4.56 & 5.14 & 4.19 & 2.62 \\
	DF Tau A & 2.69 & 13.30 & 7.21 & 7.37 \\
	DM Tau & 6.12 & 19.55 & 7.68 & 37.95 \\
	GM Aur & 3.84 & 6.72 & 1.47 & 1.74 \\
	HN Tau A & 41.71 & 63.16 & 13.52 & 35.81 \\
	LkCa15 & 111.03 & 103.30 & 14.14 & 151.65 \\
	RECX 11 & 2.40 & 9.48 & 1.09 & 0.93 \\
	RECX 15 & 42.45 & 90.01 & 13.98 & 63.32 \\
	SU Aur & 25.73 & 39.31 & 13.24 & 21.07 \\
	TW Hya & 2.64 & 3.29 & 3.63 & 2.15 \\
	UX Tau A & 104.69 & 124.23 & 13.14 & 123.16 \\
	V4046 Sgr & 12.82 & 13.09 & 5.93 & 2.86 
\enddata
\tablenotetext{}{All model-to-data reduced-$\chi^2$ statistics for simultaneous emission line fitting, transitioning from excited state [$v'$,$J'$]. All $\chi^2$ statistics are calculated between $v_{obs}$ = [-250, 250] km/s. The largest source of errors in the $\chi^2$ statistics come from the linear estimation of the FUV background continuum beneath the emission line. Because the models do not attempt to find the background continuum levels beneath each emission line, extraction of the FUV continuum had to be done manually. Targets with lower signal-to-noise have more uncertainty in the continuum flux, so the $\chi^2$ statistics become large as the errors in the continuum dominate the fitting. Only the [1,4] progression show decent fits for all targets (with $\chi^2 <$ 15), so we focus on the relative results of the [1,4] progression emission lines for the remainder of the Discussion section.}
\end{deluxetable}
\begin{figure}[t!]
	\includegraphics[angle=90, width=0.45\textwidth]{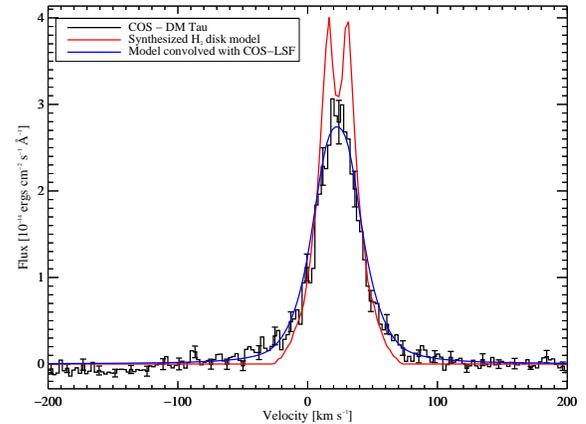}
	\caption{An example of a modeled emission line fit over a \textit{HST}-COS emission line. The target is DM Tau, one of the transitional disk targets, for emission line $\lambda$1489.57 {\AA}, which fluoresces from the Lyman band energy level [1,4]. The black line represents the observed H$_2$ fluorescent emission feature, which includes error bars every 5 bins. Each emission line observed has an intrinsic background continuum from the stellar source (see \citealt{France+14}), so this continuum was subtracted from the line before model comparisons were made to the observations. The red line is the modeled emission of 1489.57{\AA} from the DM Tau disk model. The blue line is the convolution of the modeled emission line with the COS LSF. This procedure was applied to all modeled emission lines for all targets when comparing the modeled data with FUV observations. The reduced-$\chi^2$ was calculated after the model emission lines were convolved with the COS LSF.}
	\label{fig:obs_model_convol_ex}
\end{figure}
\indent Assuming an absorption coefficient $\kappa_{\lambda} \left(r,z \right) = \sigma_{H_2} n_{[v,J]}(r,z)$, the optical depth of H$_2$ in ground state [$v$,$J$]  is described as:
\begin{equation}
	\tau_{\lambda} \left(r,z \right) = \sum_{z}^{z - H_p} z \kappa_{\lambda} \left(r,z \right)  .
	\label{tau}
\end{equation}
\indent For every vertical and radial position in the disk atmosphere that we sample $\tau_{\lambda}(r,z)$, we calculate the amount of the Ly$\alpha$ radiation that will be available for absorption by the warm H$_2$. To correct for line absorption overlap of shared Ly$\alpha$ photons, we adopt an effective optical depth $\tau_{\lambda}^{'}(r,z)$ \citep{Liu+96, Wolven+97}, defined as
\begin{equation}
	\tau_{\lambda}^{'} \left(r,z \right) = \tau_{\lambda} \left(r,z \right) \frac{\tau_{\lambda} \left(r,z \right)}{\tau_{\text{all}} \left(T(r),N(r,z) \right)} ,
\end{equation}
which corrects for the absorption, scattering, and shielding of Ly$\alpha$ photons. Figure~\ref{fig:model_rt} shows a schematic of $\tau_{\lambda}(r,z)$ for [$v'$,$J'$] = [1,4] and outlines the radiative transfer process in the disk. \\
\indent We model the emission line flux of each $\lambda_{H_2}$ produced from the cascade of transitions from energy level [$v'$,$J'$] as:
\begin{equation}
	F_{\lambda_{H_2}} = \eta S_{\lambda} \left(r,z \right) B_{mn} \sum^{\tau_{\lambda}^{'}}  \left(1 -   e^{-\tau_{\lambda}^{'} \left(r,z \right)} \right) ,
	\label{FH2}
\end{equation}
%
%
\begin{figure*}[t!]
	\includegraphics[angle=90, width=1.0\textwidth]{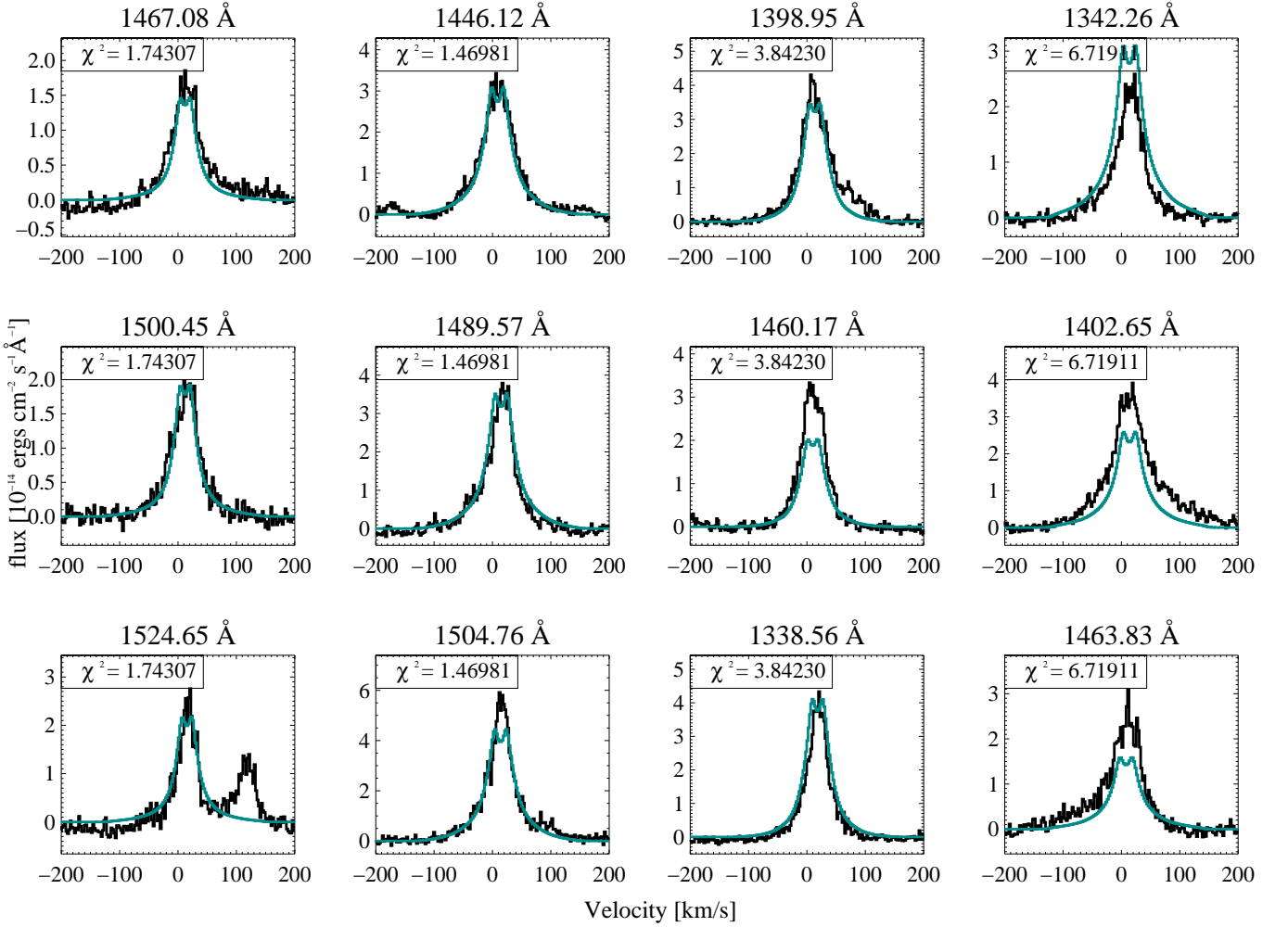}
	\caption{The resulting model and data fits of the minimum simultaneous progression $\chi^2$ statistic for GM Aur. Each column represents transitions from a common excited energy level [$v'$,$J'$]. From left to right: the left column - [$v'$,$J'$] = [1,7]; the middle-left column - [1,4]-; the middle-right column - [0,1]; the right column - [0,2]. All reduced-$\chi ^2$ values for each progression were calculated by simultaneously fitting each observed emission line profile to those estimated in a given model parameter set. The minimum reduced-$\chi ^2$ for each progression is assumed to best represent the H$_2$ fluorescence distribution. The $\chi^2$ shown in the top left of each emission line box represents the fitting of all emission lines from a given progression with one set of model parameters. Most of the observed emission lines for all targets have single-peaked line profiles (see \citealt{Brown+13}), but all the best-fit modeled emission lines show a ``double horned'' profile signature to Keplerian motions. \citet{Pontoppidan+11} points out that the single-peaked profile behavior is suggestive of a low velocity ($<$ 3 km/s) molecular wind located within a few AU of the central star and are typically modeled with an azimuthal velocity vector that is slow relative to Keplerian motion (also see \citealt{Bast+11}). Since our models leave out the azimuthal velocity component of the H$_2$ disk gas motions, it is expected that our resulting emission line profiles do not reproduce the line cores of the fluorescent features.}
	\label{fig:obs_model_progression_GMAur}
\end{figure*}
where $\eta$ represents the coverage fraction of H$_2$ in the Ly$\alpha$ radiation field \citep{Herczeg+04}, $B_{mn}$ is the branching ratio describing the  fraction of H$_2$ decaying via a given transition to ground state [$v''$,$J''$] over the whole suite of transitions available from the progression, and the source function ($S_{\lambda}(r,z)$) is defined as the Ly$\alpha$ emission line flux with wavelength $\lambda_{Ly\alpha}$, $F_{\text{Ly}\alpha}(r,z)$. \\
\indent We calculate how $F_{\text{Ly}\alpha}(r,z)$ changes as a function of radial position in the disk. Assuming that the accretion-generated Ly$\alpha$ flux originates at the stellar surface, we express the ratio of the original $F_{Ly \alpha,\star}$ to the flux the warm H$_2$ disk atmosphere receives at $r$,
\begin{equation}
	F_{Ly \alpha} = F_{\star,Ly \alpha} \frac{R_{\star}^2}{r^2} .
\end{equation}
\indent To correctly incorporate the Ly$\alpha$ radiation field, we use reconstructued stellar Ly$\alpha$ profiles created by \citet{Schindhelm+12b} and \citet{France+14}, which describe the stellar-Ly$\alpha$ flux seen by the disk surface of each target. 
After calculating the FUV H$_2$ fluorescence flux at each disk grid point in our model, we radiate the H$_2$ emission isotropically, some fraction of which is intercepted by the observer. We calculate the distance of each gas parcel radiating in the disk from the observer $s(r,z)$, based on radial and angular positions of the disk gas parcel, distance to the target, and disk inclination angle. The final modeled emission line flux produced for a fluorescence transition of H$_2$ is expressed as:
\begin{equation}
\begin{split}
	F_{\lambda_{H_2}} = \eta & F_{\star,Ly \alpha} \left(\frac{R_{\star}^2}{r^2} \right) \left(\frac{(d \cos i_{disk})^2}{s(r,z)^2} \right) \times \\
	                    & B_{mn} \sum^{\tau_{\lambda}^{'}} \left(1 -   e^{-\tau_{\lambda}^{'} \left(r,z \right)} \right)
	\label{H2flux}
\end{split}
\end{equation}
\\
\indent Using a total of 6 parameters to represent the physical conditions of the warm, ground-state H$_2$ populations in the disk atmosphere (\textit{z/r, $\gamma, q, T_{1 AU}, r_{char}, M_{H_2}$}), Equation~\ref{H2flux} characterizes the resulting emission line profiles from H$_2$ radiating from the disk. All free parameters were allowed to vary over a rough grid of controlled values to create a data cube representing the density distributions, temperature profiles, and radial radiation fields of inner disk H$_2$ around a given stellar target; see Table~\ref{tab:parm_values} for the full list of parameters explored in this study. The resulting models simulate the emission profiles produced for a given fluorescence transition $\lambda_{H_2}$, with emission flux as a function of orbital velocity. The radial velocity component of the emission line is determined by $v_{\phi}(r)$ of the emitting gas at a given radius in the disk, projected into the sight line of the observer. This model framework was used to describe the observed velocity field of single and binary systems, both close-in and extended. We caution the reader regarding the results of the close-in binary systems (e.g. V4046 Sgr), as the binary potential affects the inner disk velocity-radial relationship differently than a point mass. Therefore, the innermost H$_2$ modeled for these close-in binary systems may not be accurate, but the outer disk emission distributions will remain unaffected. \\
\indent Synthesized spectra of each H$_2$ emission line are compared to \textit{HST} observations. Each model is convolved with either the \textit{HST}-COS line spread function (LSF) (\citealt{Kriss+11}) or a normalized Gaussian distribution with FWHM characterized by the STIS E140M mode spectral resolving power (R$\sim$25,000 for TW Hya; see \citealt{Herczeg+06} for more information) prior to comparison with the observed emission line profiles. The FUV continuum level is estimated around each emission feature with a linear fit to the \textit{HST}-COS data, which is subtracted from the observations before model-to-data comparisons are made. An example of an H$_2$ emission line, with native and convolved models laid over the \textit{HST}-COS observed emission line, is shown in Figure~\ref{fig:obs_model_convol_ex}. 
%
%
%
%
%
%
%
\section{Analysis}
\begin{figure}[t]
	\includegraphics[angle=90, width=0.48\textwidth]{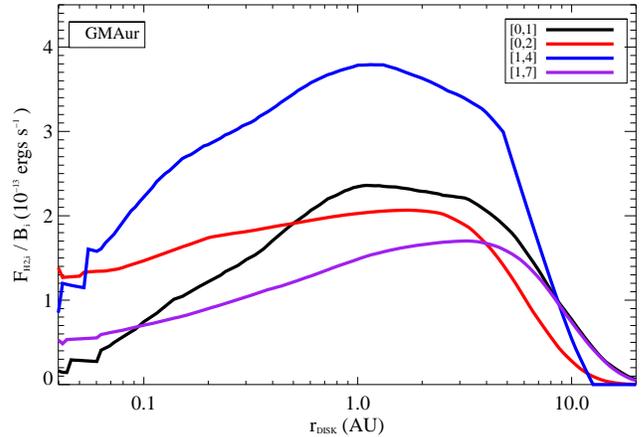}
	\caption{Using the best-fit progression model for GM Aur, we use Equation 12 (integrated over $B_{mn}$, which represents the total H$_2$ flux produced from each progression) to reproduce the observed spectrum. Each progression peaks at different radii, but the overall shape and radial extent of the distributions indicate that the bulk of the radiation for all progressions originates within the same disk annuli.}
	\label{fig:radiation_progression_GMAur}
\end{figure}

The goal of the model-to-data comparison is to find the combination of model parameters that best reproduce the observed fluorescent emission line profiles that cascade from the same excited state [$v'$,$J'$]. A reduced-$\chi ^2$ statistic is computed when comparing the observed FUV H$_2$ emission features to the entire data cube of models created for a target. We analyze the reduced-$\chi$$^2$ statistic data cube for three cases when comparing the modeled emission lines to the observations: (1) fitting individual emission lines; (2) simultaneously fitting all H$_2$ emission lines fluorescing from the same excited energy level [$v'$,$J'$]; (3) fitting only the red wings of the emission lines. The first point was used to set the initial range of temperature and density model parameters of warm H$_2$ in each disk surface. The third was explored to mitigate the potential influence of a warm molecular wind component that was unresolved at the spectral resolving power of \textit{HST}-COS. The results of (3) proved inconclusive, which found no significant differences between the red and blue wing line shapes, suggesting that the models are not sensitive to an unresolved warm H$_2$ disk wind. We focus on the results of (2), which best describe the generalized behavior of the warm H$_2$ disk atmosphere populations. We simultaneously fit 3 observed fluorescent H$_2$ transitions for each progression as the most representative of the H$_2$ radiation distributions in each PPD. \\
\indent Table~\ref{tab:min_chi2} shows the minimum reduced-$\chi ^2$ statistics for all targets when simultaneously fitting the 3 progression emission lines from excited state [$v'$,$J'$].  Not all minimum reduced-$\chi ^2$ simultaneous progression fits for [0,1] and [0,2] were ``good'', however (i.e., some sources displayed reduced-$\chi ^2$ $>$ 25). 
Many of the strongest lines from [0,1] and [0,2] share similar $\lambda_{H_2}$, which makes complex line profiles that depend on the shape of the stellar-Ly$\alpha$ profile illuminating the warm H$_2$ disk populations to these excited states. The [1,7] and [1,4] progressions are more reliable tracers of the warm H$_2$ disk atmosphere, and the brightest emission lines in our survey cascade from the [1,4] progression. For the vast majority of the targets, the largest stellar Ly$\alpha$ fluxes pump the warm H$_2$ disk populations to the [1,4] energy level. This makes the line profile flux fitting more accurate for the [1,4] progression, providing the overall best model fits to the observe FUV emission. \\
\indent We will focus our discussion around the inner disk diagnostics of the best-fit [1,4] progression for all targets. This progression has good reduced-$\chi ^2$ fits ($\leq$ 15) and by-eye model-to-data comparisons for every target in our survey. Figure \ref{fig:obs_model_progression_GMAur} shows an example of minimum reduced-$\chi ^2$ modeled progression lines to those observed with \textit{HST}-COS for GM Aur. Figure \ref{fig:radiation_progression_GMAur} presents the resulting radial radiation distribution for each best-fit progression for GM Aur. While each progression peaks at somewhat different radii, the majority of the radiation distributions originate within similar annuli of the disk. This behavior is typical for all PPD targets that have good minimum reduced-$\chi ^2$ fits for all or most progressions. 
%
%
\begin{figure*}[htp!]
	\includegraphics[angle=90, width=1.1\textwidth]{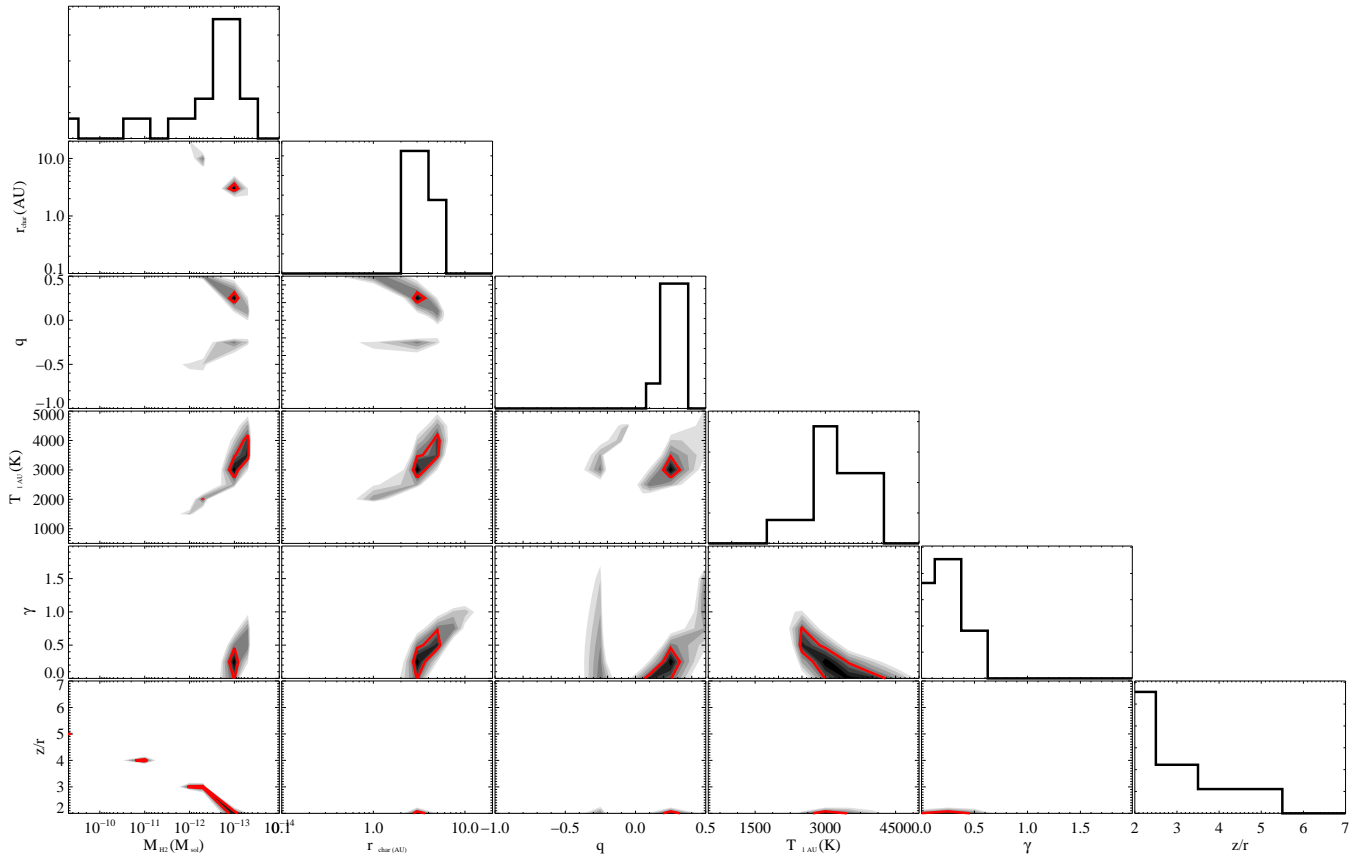}
	\caption{Marginalized distributions of the [1,4] progression reduced-$\chi ^2$ fits for RECX-11. The uncertainties in the best-fit model parameters are measured as the range of values that encompass 68\% of the distribution area and are highlighted in red contour outlines. The posterior marginalized distributions for each parameter against all other model parameters are shown as the 2D plots at the top of each column. A Gaussian distribution was estimated for each posterior distribution, and final uncertainty estimates for each model was calculated as the FWHM of the distribution. These errors were later used to estimate errors in radial radiation distribution boundaries.}
	\label{fig:chisqr_contours_RECX11}
\end{figure*}
\subsection{Uncertainty Estimation and Parameter Degeneracies}
\indent Errors in each best-fit parameter per progression are determined after marginalizing the minimum reduced-$\chi ^2$ parameter fits over all free parameters. Uncertainties are measured as the range of values that encompass 68\% of the distribution area, representing the 1-$\sigma$ uncertainties for a Gaussian distribution. The modeled parameter space was crudely varied over a large range of values for each free variable, so a Gaussian distribution was fit over each marginalized best-fit parameter uncertainty space, and the FWHM of each Gaussian fit was calculated as the uncertainty in each model parameter. \\ 
%
\indent Figure \ref{fig:chisqr_contours_RECX11} displays the reduced-$\chi ^2$ marginalized parametric space for each variable in our modeling framework, with filled contours representing the 2-$\sigma$ uncertainty in the parameter space. Since each parametrized uncertainty is taken within the 1-$\sigma$ error contours of each marginalized distribution, the uncertainties outlined in red represent the 1-$\sigma$ errors in the model parameters. \\
\indent There are noticeable degeneracies amongst several of the parameters; for example, the total mass of emitting H$_2$ and vertical position of the disk atmosphere ($M_{H_2}$, $z/r$) show a trend that requires more mass contributing to the emission lines as the disk height above the disk mid plane increases. This trend makes sense - to produce the same amount of flux in the modeled emission lines, the total mass of H$_2$ contributing to the emission must increase as the density of H$_2$ decreases with vertical disk height above the mid plane. The optical depth of the disk atmosphere must remain the same to output the same observed emission line flux, and this relationship between the free parameters maintains the required optical depth. What is important to note is that the models produced are used as a means to describing the H$_2$ emission flux arising from the inner disk atmosphere. Despite the degeneracies in several parameter pairings relating to the total flux, the radiation distribution of H$_2$ emission is unaffected by these degeneracies. \\
\indent We note that our choice in using the reconstructed stellar LyA flux incident on the disk from \citet{Schindhelm+12b} may exacerbate degeneracies in the disk parameters. The \citet{Schindhelm+12b} reconstructed Ly$\alpha$ profiles rely on the same H$_2$ emission features explored in this study, but we remind the reader that the stellar Ly$\alpha$ flux incident on the H$_2$ disk scales with the re-emitted H$_2$ flux (see Eqn~\ref{H2flux}) and has no effect on the modeled distribution of H$_2$ flux in each disk. The disk parameters may respond to an inaccurate Ly$\alpha$ flux, but the degeneracies in the disk parameters (for example, the response of $M_{H_2}$ and $z/r$ to the total H$_2$ flux) would scale to best describe the H$_2$ radiation that recreate the observed emission profiles. Therefore, the reconstructed Ly$\alpha$ profiles will not change the radial behavior of the best-fit H$_2$ flux models.
%
\subsection{The Radial Extent of H$_2$ Emission}
\begin{figure}[t]
	\includegraphics[angle=90, width=0.48\textwidth]{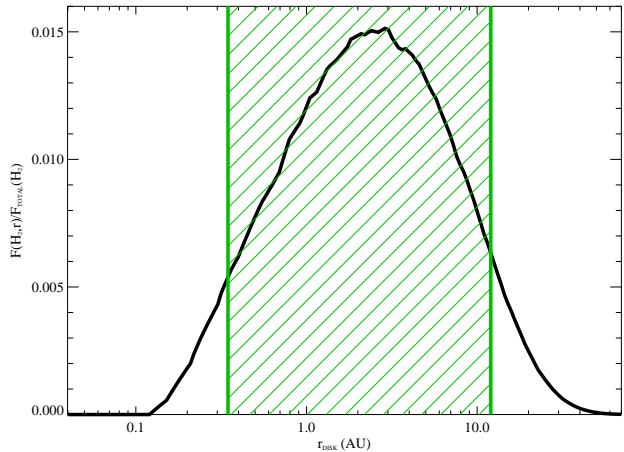}
	\caption{The inner and outer radial boundaries which define where 90\% of the total radiation is arising from the disk. The black line represents the normalized radial distribution of the [1,4] progression emission to the total amount of flux produced by the [1,4] progression for SU Aur. The green vertical lines show the radial boundaries that encapsulate 90\% of the total emission. For r$_{in}$, we start at the outermost radius and integrate inward to smaller radii the disk until 95\% of the total [1,4] progression flux is accounted for. Likewise, r$_{out}$ is defined by starting at the innermost radius defined in our models and integrating the progression flux out until 95\% of the total emission flux is accounted for. The resulting annulus between r$_{in}$ and r$_{out}$ represents the ring of disk the majority of the observed FUV H$_2$ fluorescent emission originates.}
	\label{fig:rin_rout_SUAur}
\end{figure}

\indent 
Figure \ref{fig:radiation_progression_GMAur} presents an example of the radiation distributions of H$_2$ fluorescence flux produced in the disk for each progression explored in this study. We focus our analysis on the [1,4] radiation distributions for all targets in our survey to define inner and outer radial disk boundaries, which describe where the bulk (90\%) of the emitting H$_2$ atmosphere resides. We define the 90\% emitting region as follows:
%
\begin{equation*}
	F_{H_{2},obs} = \begin{cases}
		  \frac{F(H_2,r)}{F_{tot}(H_2)} & \leq 0.95 \text{ for } r > r_{in} \\
			\frac{F(H_2,r)}{F_{tot}(H_2)} & \leq 0.95 \text{ for } r < r_{out}
									 \end{cases}
\end{equation*}
\indent We use r$_{in}$ and r$_{out}$ to evaluate the evolutionary behavior of the H$_2$ radiation. Figure \ref{fig:rin_rout_SUAur} presents a schematic of how the inner and outer radial boundaries encapsulate 90\% of the total H$_2$ flux produced in the disk atmosphere. We analyze potential evolutionary characteristics of the molecular disk atmosphere by comparing the FUV H$_2$ radiation distributions to other dust and molecular disk observables.
\begin{figure}[htp]
	\includegraphics[angle=90, width=0.48\textwidth]{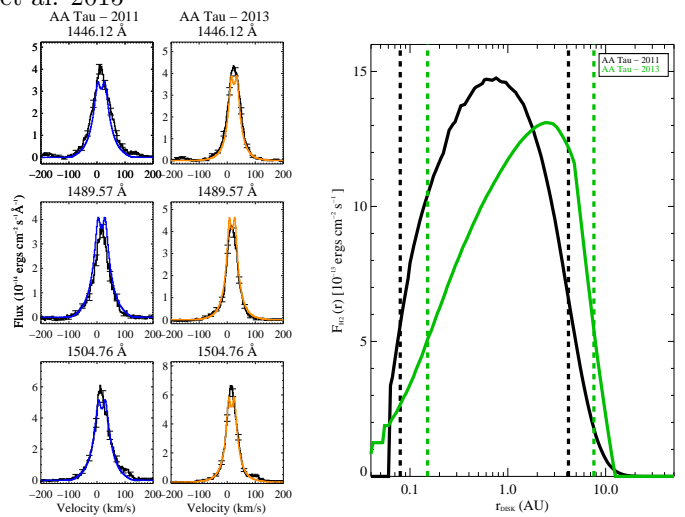}
	\caption{We present a comparison of the [1,4] progression observed with \textit{HST}-COS for AA Tau during the 2011 and 2013 epochs. On the left: The column under the 2011 label represent the 3 observed fluorescent emission line profiles cascading from the [1,4] excited state, with overlaid best-fit modeled emission lines in blue. The column to the right, labeled 2013, shows the observed [1,4] fluorescent emission lines, with modeled emission lines overlaid in orange. The 2013 observed emission line profiles appear narrower than their 2011 counterparts \citep{Schneider+15}. On the right: the comparison of the total [1,4] progression flux radiating from the disk of AA Tau in the 2011 and 2013 observations. The 2013 models predict that the observed H$_2$ fluorescence emission  originates from further out in the disk (the peak of the radiation located at r$_{peak}$ = 2.50 AU) than the 2011 radiation distribution (r$_{peak}$ = 0.75 AU), a consequence of the inner disk ``shadowing'' produced by the extra absorber on the AA Tau sightline \citep{Bouvier+13}.}
	\label{fig:AATau}
\end{figure}
%
%
%
%
%
\subsection{Case Study: Model Robustness using AA Tau}
We explore how robust our modeling framework is at identifying where the fluorescing H$_2$ resides in PPDs. We compare two epochs of \textit{HST}-COS data on AA Tau (2011 and 2013), where the 2013 observations occur during a ``dimming'' event from X-ray to near-IR wavelengths. 
Based on the duration of the dimming, \citet{Bouvier+13} suggest an obscuration at r $>$ 8 AU; this hypothesis is strengthened by the gas-to-dust ratio (N$_H$/A$_v$) of the absorber and the evolution of the FUV H$_2$ emission \citep{Schneider+15}. We utilize the line profile changes between AA Tau \textit{HST}-COS FUV observing epochs to determine how those changes relate to radial H$_2$ radiation distributions in the disk. \\
\indent There are noticeable differences between the observed FUV H$_2$ emission line profiles of the 2011 and 2013 AA Tau epochs. The 2013 emission lines are narrower with slightly larger peak fluxes than the same H$_2$ emission lines observed in 2011 \citep{Schneider+15}. This suggests that less flux is contributed from the innermost disk. The modeling results for the [1,4] progression are shown in Figure \ref{fig:AATau}. Each AA Tau epoch was modeled independently, and the models reproduce the same rest wavelength emission lines. Figure \ref{fig:AATau} also shows the radiation distributions of [1,4] fluorescence for each epoch in the AA Tau disk. The 2011 emission includes a large contribution from material inside 1 AU (r$_{in,2011}$ = 0.08$\pm$0.01 AU; r$_{peak,2011}$ = 0.75 AU; r$_{out,2011}$ = 4.17$\pm$2.04 AU), while the 2013 [1,4] emission ``appears'' to have shifted outward in the disk (r$_{in,2013}$ = 0.15$\pm$0.02 AU; r$_{peak,2013}$ = 2.50 AU; r$_{out,2013}$ = 7.59$\pm$2.75 AU). Our models indicate that the inner radius of detectable H$_2$ fluorescence from the [1,4] progression has moved outward radially in the disk as the ``extra absorber'' moved into our field of view in the AA Tau disk. \citet{Schneider+15}, using an independent modeling technique to estimate the radial origins of H$_2$ fluorescence in the AA Tau disk, come to a similar conclusion: the observed 2013 H$_2$ emission within $\sim$ 1 AU is reduced compared to 2011. Additionally, \citet{Schneider+15} find that the outer radial extent of the H$_2$ fluorescence luminosity doesn't change significantly between epochs, which is a result consistent within the errors on our modeled r$_{out}$ estimates of the AA Tau epochs. \\
\begin{figure*}[htp]
	\includegraphics[angle=90, width=1.0\textwidth]{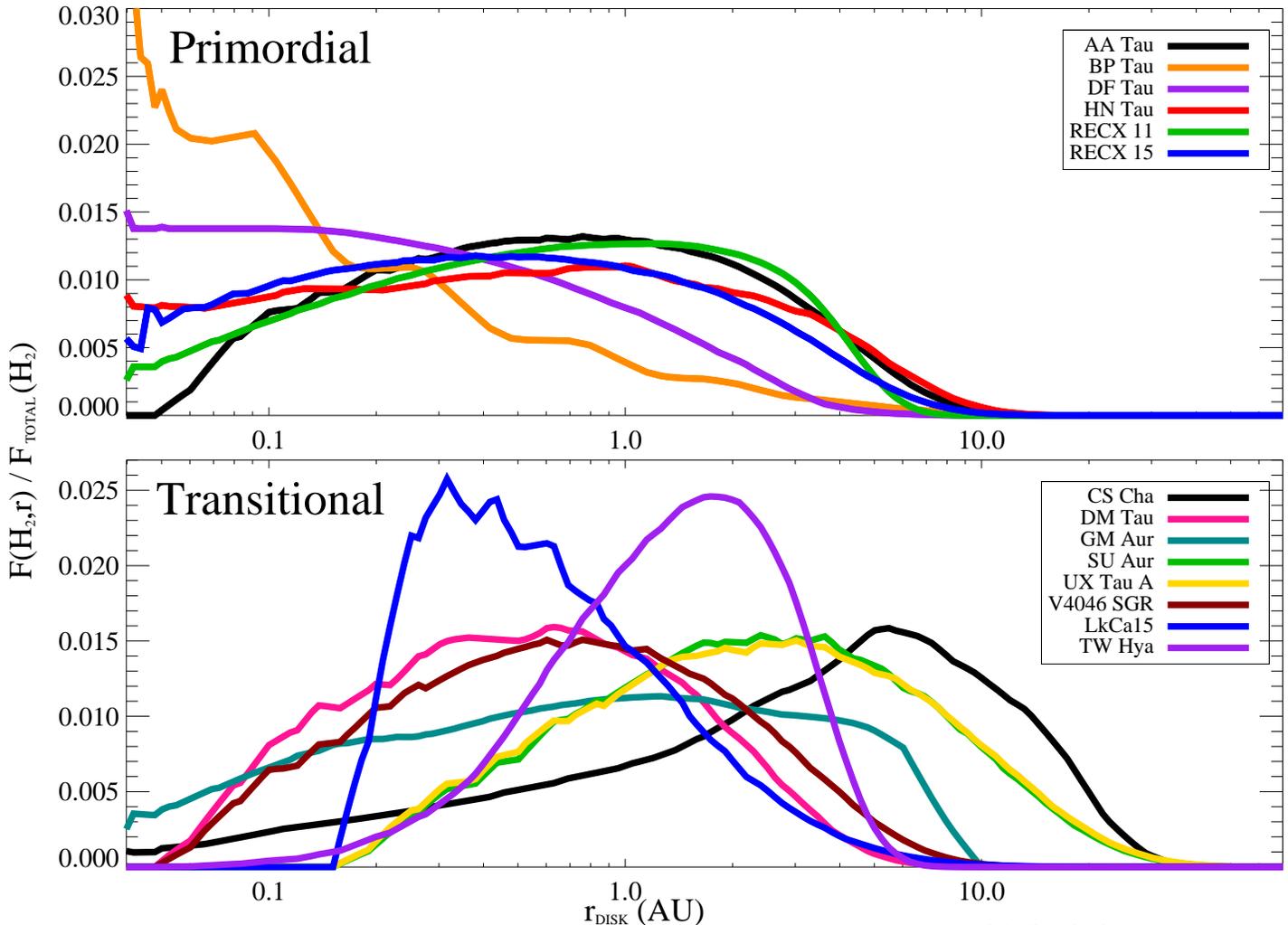}
	\caption{The normalized modeled radiation field distribution of H$_2$ fluorescence cascading from the [$v'$,$J'$] = [1,4] energy level for all targets. Each radiation distribution was calculated from the minimum reduced-$\chi ^2$ model parameters that best reproduce the observed H$_2$ emission lines. The top plot represents radiation distributions for all primordial disk targets, and the bottom plot shows the distributions for transition disk objects. The two disk evolution types appear to show an evolving H$_2$ FUV radiation field; primordial disks generally radiating more inward in the disk, with the bulk of the radiation occurring within r $\lesssim$ 1 AU, and transition disk H$_2$ radiation starting at larger radii (r $\sim$ 0.1 AU) extending to larger radii (r $\sim$ 10 AU).}
	\label{fig:radiation_distributions}
\end{figure*}
\indent The ``extra absorber'' obscures the inner disk H$_2$ fluorescence in the 2013 \textit{HST}-COS FUV spectrum, making AA Tau appear as a disk with a deficit of inner disk emission - effectively, a pseudo-transition disk. Our modeling framework was capable of identifying the change in emission line profiles between the 2011 and 2013 AA Tau observations and found that the bulk of the 2013 AA Tau [1,4] radiation in the disk originated at larger radii than the 2011 H$_2$ fluorescence. We expect our models are therefore capable of distinguishing between H$_2$ fluorescence evolution in differing disk types.
\section{Discussion}
We created 2D radiative transfer models to simulate observed \textit{HST}-COS and -STIS FUV H$_2$ emission lines to understand where the majority of the radiation arises in PPDs. We use the best-fit model results to define the inner and outer radii of warm H$_2$(r$_{in}$, r$_{out}$) and examine if and how the molecular distributions change as PPDs evolve. We compare r$_{in}$ and r$_{out}$ to other dust and molecular tracers that help describe the evolutionary state of the PPDs. Table \ref{tab:disk_parms} provides a detailed list of inner disk observables for each target, including dust cavity radius (r$_{cavity}$) and inner disk CO radius (r$_{in,CO}$). We also look at where the theoretical snow lines in the disks exist and how these radii relate to the H$_2$ disk emission. 
%
%
%
%
\subsection{Radiation Distribution of Modeled H$_2$ Fluorescent Emission}
\begin{figure}[t!]
	\includegraphics[angle=90, width=0.48\textwidth]{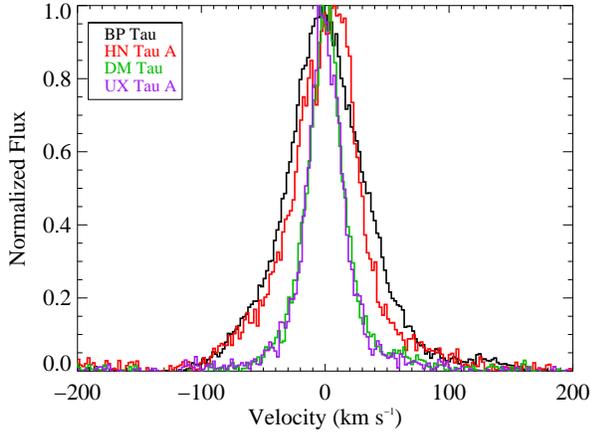}
	\caption{A comparison of observed [1,4] progression line profiles of targets with inclination angles between 30$^{\circ}$ and 40$^{\circ}$. The two broadest line profiles, BP Tau and HN Tau A, are in the primordial phase. The two narrowest line profiles, DM Tau and UX Tau A, are in the transition phase.}
	\label{fig:prim_v_trans}
\end{figure}

\begin{figure}[htp]
	\includegraphics[angle=90, width=0.48\textwidth]{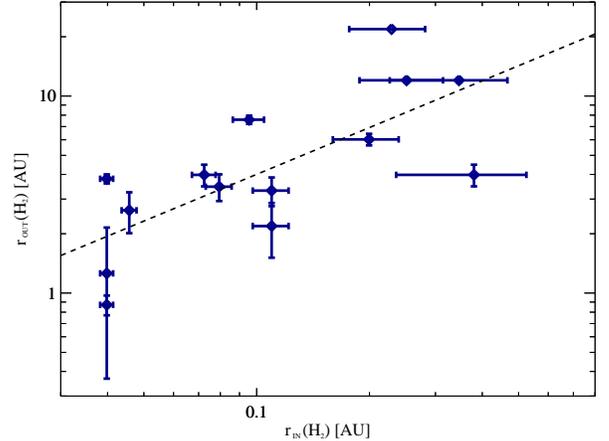}
	\caption{We present the relation between the estimated r$_{in}$ and r$_{out}$ quantities, determined from the best-fit modeled radiation distributions for all targets. The inner radial boundary (r$_{in}$) is defined as the inner radius of H$_2$ radiation in the disk that marks where at least 90\% of the total radiation is accounted for in the outer disk. Likewise, the outer radial boundary (r$_{out}$) is defined as the outermost radius of H$_2$ radiation that encompasses 90\% of the total amount of radiation accounted for in the inner disk. The blue diamonds with error bars  represent each modeled r$_{in}$ and r$_{out}$, and the black dashed line represents a linear fit to the data. The Spearman rank correlation coefficient ([$\rho$, $n$] = [0.70, 5.5 $\times~10^{-3}$]) between the two radial quantities suggest a strong increasing trend between them, indicating that the whole emitting region is moving outward.}
	\label{fig:rin_v_rout}
\end{figure}

Figure \ref{fig:radiation_distributions} presents the normalized radial distributions of warm H$_2$ transitioning from excited state [1,4] for all targets. We modeled 6 primordial disks (AA Tau, BP Tau, DF Tau A, HN Tau A, RECX-11, and RECX-15) and 8 transition disks (CS Cha, DM Tau, GM Aur, LkCa 15, SU Aur, TW Hya, UX Tau A, and V4046 Sgr) to compare the radial distribution of warm H$_2$ in the disk atmospheres as the dust disk evolves. The H$_2$ radial distributions of the different dust disk stages appear qualitatively different. The primordial disk population (top plot in Figure \ref{fig:radiation_distributions}) generally starts radiating significantly in the very inner disk (r $\lesssim$ 0.05 AU), and the radiation only extends out to a few AU, consistent with the simple estimates of the average H$_2$ emitting radius presented by \citet{France+12b}. The generalized transition disk radiation behavior (bottom plot) starts further out in the disk (r $\sim$ 0.1 AU) and extend significantly further out into the disk (r $\sim$ 10 AU). These different behaviors suggest structural changes in any of all of the following: the spatial distributions of warm H$_2$ in populations [$v$,$J$]; the degree of Ly$\alpha$ penetration into the disk by clearing H$_2$ from the inner disk atmosphere; or the evolution of the disk surface temperature distribution. This evolving radiation structure is also observable in the line profiles of the [1,4] progression, as seen in Figure~\ref{fig:prim_v_trans}. As the PPDs in our survey evolve from primordial to transition disks, the majority of the observed H$_2$ emission migrates to larger radii. \\
%
\begin{deluxetable*}{l c c c c c c c l}\tablenum{5}
\tabletypesize{\footnotesize}
\tablecaption{Disk Parameters from Results \& Literature
	\label{tab:disk_parms}}
\tablewidth{0pt}
\tablecolumns{9}
\tablehead{
	\colhead{Target} & \colhead{n$_{13-31}$} & \colhead{$\dot{M}$\tablenotemark{a}} &
	\colhead{r$_{in,H_2}$} & \colhead{r$_{out,H_2}$} & \colhead{r$_{in,CO}$} & 
	\colhead{r$_{cavity}$} & \colhead{T(H$_2$)} & \colhead{ref.\tablenotemark{b}} \\	
	            {}     &                {}              & {(10$^{-8}$ $M_{\sun}$ yr$^{-1}$)}   & 
							{(AU)}    & {(AU)}                  & {(AU)}               &
							{(AU)}     &   {K}  & }
\startdata

	AA Tau & -0.51 & 1.5 & 0.08 $\pm$ 0.01 & 3.47 $\pm$ 0.54 & 0.10 & ... & 4000$^{+250}_{-1500}$  & 2,11 \\
	BP Tau & -0.58 & 2.9 & 0.04 $\pm$ 0.01 & 0.87 $\pm$ 0.10 & 0.03 & ... & 2000$\pm$300 & 1,2,11 \\
	CS Cha & 2.89 & 5.3 & 0.23 $\pm$ 0.05 & 21.88 $\pm$ 4.68 & ... & 40 & 2500$\pm$400 & 5 \\
	DF Tau A & -1.09 & 17.7 & 0.04 $\pm$ 0.01 & 1.26 $\pm$ 0.89 & 0.10 & ... & 1500$^{+1000}_{-100}$ & 11 \\
	DM Tau & 1.30 & 0.29 & 0.11 $\pm$ 0.01 & 2.19 $\pm$ 1.48 & ... & 3 & 2000$\pm$500 & 3,4,10 \\
	GM Aur & 1.76 & 0.96 & 0.10 $\pm$ 0.01 & 7.59 $\pm$ 2.75 & 0.20 & 20 & 3000$\pm$450 & 1,2,3,4,11 \\
	HN Tau A & -0.44 & 0.13 & 0.04 $\pm$ 0.01 & 3.80 $\pm$ 0.20 & ... & ... & 2500$\pm$750 &  \\
	LkCa 15 & 0.62 & 0.31 & 0.20 $\pm$ 0.04 & 6.03 $\pm$ 2.45 & 0.10 & 46 & 1500$^{+1250}_{-200}$ & 1,3,6,10,11 \\
	RECX-11 & -0.80\tablenotemark{c} & 0.03 & 0.07 $\pm$ 0.02 & 3.98 $\pm$ 2.00 & ... & ... & 3000$^{+1000}_{-1250}$ & \\
	RECX-15 & -0.20\tablenotemark{c} & 0.10 & 0.05 $\pm$ 0.01 & 2.63 $\pm$ 1.62 & ... & 7.5 $\pm$ 1.5 & 2500$\pm$350 &  12 \\
	SU Aur & 0.74 & 0.45 & 0.35 $\pm$ 0.12 & 12.02 $\pm$ 3.47 & ... & ... & 1500$^{+1250}_{-300}$ & \\
	TW Hya & 0.20\tablenotemark{c} & 0.02 & 0.38 $\pm$ 0.14 & 3.98 $\pm$ 1.0 & 0.1$^{+0.2}_{-0.04}$ & 4 & 2000$^{+500}_{-150}$ & 3,4,10 \\
	UX Tau A & 1.83 & 1.00 & 0.25 $\pm$ 0.06 & 12.03 $\pm$ 3.46 & 0.30 & 25 & 1500$^{+1000}_{-300}$ & 3,6,11 \\
	V4046 Sgr & 0.32\tablenotemark{c} & 1.30 & 0.11 $\pm$ 0.01 & 3.31 $\pm$ 1.82 & ... & 14 & 2000$\pm$500 & 8,9 \\
	
\enddata
	
	\tablenotetext{a}{~All $\dot{M}$ values taken from \citet{Ingleby+13}.}
	\tablenotetext{b}{~(1) \citet{Akeson+05}; (2) \citet{Andrews+Williams+07}; 
	                   (3) \citet{Andrews+11}; (4) \citet{Calvet+05}; (5) \citet{Espaillat+07a};	
	                   (6) \citet{Espaillat+07b}; (7) \citet{France+12b}; (8) \citet{Rapson+15}; 
										 (9) \citet{Rosenfeld+13}; (10) \citet{Salyk+09}; (11) \citet{Salyk+11b}; 
										 (12) \citet{Woitke+11} } 
  \tablenotetext{c}{~For n$_{13-31} \mu m$ values not listed in \citet{Furlan+09}, we use Equation~\ref{eqn:n1331} to estimate the observable from known or modeled dust SED.}
\end{deluxetable*}
\indent We compare estimates of r$_{in}$ and r$_{out}$ to investigate the idea that the radial distributions of fluorescing H$_2$ migrate outward in the disks as PPDs evolve. Figure \ref{fig:rin_v_rout} presents a comparison of r$_{in}$ and r$_{out}$, which shows the annulus of H$_2$ emission extending further out into the disk as the inner disk radius moves outward. A line can be fit to represent the relationship between the inner and outer radiating disk radii for our survey targets:
\begin{equation}
	\text{log}_{10}(r_{out}(H_2)) = 0.79 \text{ log}_{10}(r_{in}(H_2)) + 1.39 ,
\end{equation}
where both log$_{10}$($r_{in}$(H$_2$)) and log$_{10}$($r_{out}$(H$_2$)) are in units of AU, and the coefficients [1.39 $\pm$ 0.22, 0.79 $\pm$ 0.21] are computed from a $\chi ^2$ minimization ($\chi_{min}^2$=0.896) of a linear function between log$_{10}$($r_{in}$(H$_2$)) and log$_{10}$($r_{out}$(H$_2$)). The Spearman rank correlation coefficient between r$_{in}$ and r$_{out}$ indicates a statistically significant correlation between the variables ($\rho$ = 0.70) with a small probability that the sample is randomized ($n$ = 5.5 $\times$ 10$^{-3}$), providing additional evidence that support the migration of the radial H$_2$ emission as PPD warm dust dissipates from the inner disk.
\subsection{Comparison to Dust Evolution}
\begin{figure}[htp]
\begin{center}
	\includegraphics[angle=90, width=0.48\textwidth]{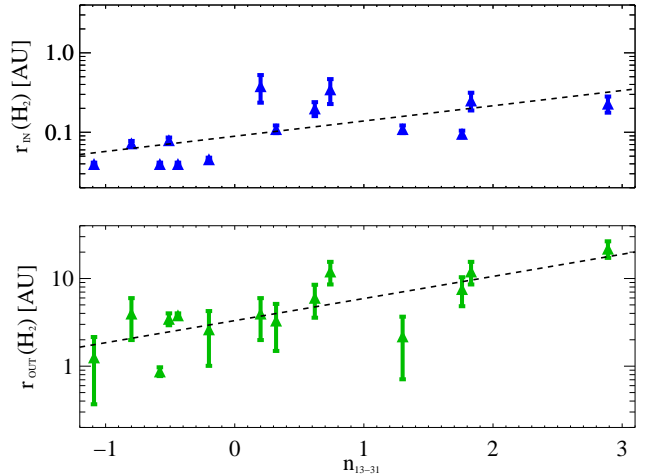}
	\caption{Comparison of r$_{in}$ and r$_{out}$ with an observable dust evolution diagnostic $n_{13-31}$ \citep{Furlan+09}. In the top plot: Each blue triangle with error bars represents each target point in our survey. The black dashed line represents the best-fit linear correlation between r$_{in}$ and n$_{13-31}$. In the bottom plot: Each green triangle with error bars represents each target point in our survey. The black dashed line represents the best-fit linear correlation between r$_{out}$ and n$_{13-31}$. In both plots, a clear increasing trend is seen in the radial H$_2$ emission boundaries as the warm dust disk content evolves.}
	\label{fig:radii_v_n1331}
\end{center}
\end{figure}
We compare results from our modeled H$_2$ [1,4] progression radial distributions with dust disk evolution diagnostics to gain insight into how the molecular inner disk environment of PPDs changes as dust grains clear. We identify PPD evolution  using observed color-color changes in the near- to mid-IR SED slope of the disk, which provides an estimate of the degree of warm dust clearing (see \citealt{Espaillat+14}). We interpret the slope of each target SED with the observable $n_{13-31}$ \citep{Furlan+09}:
\begin{equation}
	n_{13-31} = \frac{log(\lambda_{31}F_{\lambda_{31}})-log(\lambda_{13}F_{\lambda_{13}})}{log(\lambda_{31})-log(\lambda_{13})} ,
	\label{eqn:n1331}
\end{equation}
which is dominated by longer wavelength continuum emission from the optically-thick dust in the disk and is sensitive to the degree of dust settling towards the disk midplane \citep{DAlessio+06}. For many targets in this work, $n_{13-31}$ were available in \citet{Furlan+09}, but for targets not included in the \citet{Furlan+09} survey, we calculate $n_{13-31}$ with known or modeled disk SEDs (for example, an intricate model of V4046 Sgr SED was found by \citealt{Rosenfeld+13}). We interpret the results of $n_{13-31}$ as follows: if $n_{13-31}$ $<$ 0, the inner dust disk is optically thick, essentially a primordial disk; if $n_{13-31}$ $\geq$ 0, the disk dust is optically thin, indicative of dust clearing or settling and evidence for PPD evolution into the transition state \citep{Lada+87,Strom+89,Andre+Montmerle+94}. Table \ref{tab:disk_parms} provides a list of $n_{13-31}$ values for all targets in this survey. \\
\indent A comparison of the [1,4] emission boundaries (r$_{in}$, r$_{out}$) to $n_{13-31}$ is made in Figure \ref{fig:radii_v_n1331}. The top figure shows the relationship between r$_{in}$ and $n_{13-31}$, and the bottom figure shows r$_{out}$ versus $n_{13-31}$. The triangles in both plots represent each target in our survey, and the black dashed line in each plot shows the linear correlation between r$_{in}$ versus $n_{13-31}$ and r$_{out}$ versus $n_{13-31}$. It is apparent that the molecular inner and outer disk emission radii show a positive correlation with the dust disk evolution: the Spearman rank correlation coefficient for r$_{in}$ versus $n_{13-31}$ is 0.72 ($n$ = 4.0 x 10$^{-3}$), and $\rho$ = 0.69 for r$_{out}$ versus $n_{13-31}$ ($n$ = 6.9 x 10$^{-3}$). Both correlation coefficients suggest a strong increasing trend in the radial outward migration of the FUV H$_2$ radiation as the warm dust disk evolves in the disk samples. The linear correlation between r$_{in}$ and $n_{13-31}$ is expressed as:
\begin{equation}
\begin{split}
	\text{log}_{10}(r_{in}(H_2)) = & (0.19 \pm 0.07) \times n_{13-31} \\
																 & - (1.05 \pm 0.08) ,
\end{split}
\end{equation}
and the linear correlation between r$_{out}$ and $n_{13-31}$ is expressed as:
\begin{equation}
\begin{split}
	\text{log}_{10}(r_{out}(H_2)) = & (0.25 \pm 0.06) \times n_{13-31} \\
																	& + (0.52 \pm 0.07).
\end{split}
\end{equation}
\begin{figure}[htp]
	\includegraphics[angle=90, width=0.48\textwidth]{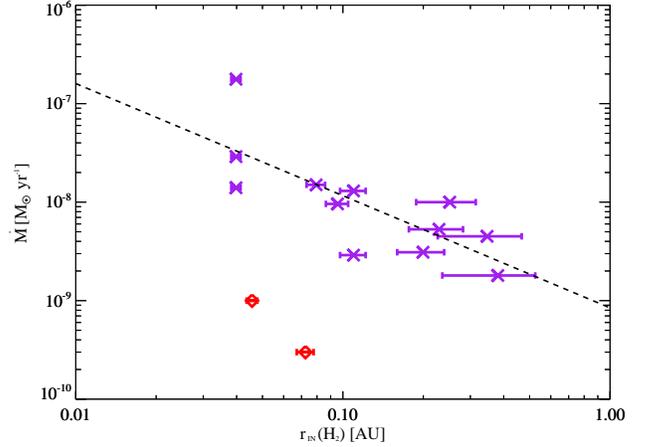}
	\caption{Comparison of the modeled inner H$_2$ emission radius to the mass accretion rate of the target (from \citealt{Ingleby+13}). The purple x-points represent all targets with mass accretion rates $>$ 10$^{-9}$ $M_{\odot}$ yr$^{-1}$, while the red diamonds represent the RECX targets (which have low mass accretion rates for primordial PPD targets). The black line is a negative correlation fit through all the purple points, suggesting that the mass accretion rate decreases as r$_{in}$ increases. Since the accretion luminosity, more specifically the stellar Ly$\alpha$ flux produced by the accretion, is directly related to the H$_2$ emission observed, it is important to note that r$_{in}$ is not necessarily correlated to the flux produced by the mass accretion rate. Instead, r$_{in}$ is sensitive to the observed emission line width, which is independent of the stellar incident flux.}
	\label{fig:rin_v_mdot}
\end{figure}
We note that, for all transition disks in this study, r$_{out}$ is found to be within the dust gap radius. One interpretation of this result, paired with the correlation between r$_{in}$ and $n_{13-31}$, is that the H$_2$ FUV radiation observed from the inner PPD atmosphere lags behind the dust disk evolution. \\
\indent This does not automatically mean that the molecular content of the disk is clearing, and we need further evidence of evolution with other inner disk molecular tracers before we can make this distinction. \citet{France+12b} outlined the conditions needed in the H$_2$ disk atmosphere to produce Ly$\alpha$-pumped H$_2$ fluorescence. The opacity of absorbing H$_2$ in ground-state [$v$,$J$] must be large, with excitation temperatures T$_{exc}$ $>$ 1500K, and the mass accretion rate ($\dot{M}$) onto the proto-star must be large enough to produce enough Ly$\alpha$ photons to stimulate the molecules. The mass accretion rate implies there is a reservoir of material in the inner regions of PPDs that feeds onto the proto-star, and a decrease in $\dot{M}$ over time (e.g., \citealt{Muzerolle+00}) strongly suggests that the inner disk material is being depleted. 
Figure \ref{fig:rin_v_mdot} shows the relationship between $\dot{M}$ and r$_{in}$(H$_2$), with purple points representing r$_{in}$(H$_2$) and $\dot{M}$ for all targets except the RECX targets, which are represented at red diamonds. All mass accretion rates are taken from \citet{Ingleby+13}. Figure \ref{fig:rin_v_mdot} shows a negative correlation between $\dot{M}$ and r$_{in}$(H$_2$), with Spearman rank correlation [$\rho$,$n$] = [-0.80, 1.9 x 10$^{-3}$] (not including the RECX targets), suggesting that the H$_2$ atmosphere may be physically thinned or in different ground-state populations not suitable for Ly$\alpha$-pumping in the very inner disk regions of evolved PPDs. The outlier points in Figure \ref{fig:rin_v_mdot}, RECX-11 and RECX-15, appear to have abnormally low mass accretion rates given the evolutionary stage of the disks \citep{Ingleby+11}, and more targets of varying evolution may be needed to understand if this result is universal among a large sampling of PPDs. It is important to note that r$_{in}$ is primarily dervied from the observed line widths of H$_2$ emission profiles, so determination of r$_{in}$ is largely independent of the incident FUV flux. \\
\indent The link between $\dot{M}$ and r$_{in}$(H$_2$) suggests that the inner disk is clearing of material as the mass accretion rate declines. 
One explanation for this correlation is that the warm H$_2$ atmosphere dissipates with the small dust grains. Dust grains present in the disk atmospheres of primordial disks may give warm H$_2$ a formation site to replenish molecules lost to photo-dissociation and stellar accretion (see \citealt{Augason+70, Habart+04, Fleming+10}). As the dust grains clear out and settle towards the disk midplane or evaporate from the inner disks of evolving PPDs, the warm H$_2$ atmosphere no longer has a formation site to maintain the molecular reservoir. Via accretion and photo-dissociative processes with FUV continuum photons between 912 - 1120 {\AA}, the leftover warm H$_2$ will continue to disperse, even as the accretion flux decreases. This leaves an optically thin ($N(H_2) \lesssim 10^{18} cm^{-2}$) path for stellar Ly$\alpha$ to reach the warm H$_2$ material at larger disk radii (r $>$ 3 AU).  \\
\indent The migration of r$_{out}$(H$_2$) with increasing n$_{13-31}$ also suggests that neutral hydrogen (HI) is being cleared from the inner disks of transitional PPDs. Photo-excitation via stellar Ly$\alpha$ drives the H$_2$ fluorescence observed in the disk atmospheres, and as the emitting H$_2$ is observed further out in the disk, there must be new paths open for stellar UV radiation to reach the outer disk material. In primordial disks, HI re-processes and scatters incident stellar Ly$\alpha$ down into the inner disk \citep{Fogel+11} while H$_2$ self-shields the radiation from penetrating to the outer disk, preventing the stellar Ly$\alpha$ from reaching the outer disk effectively. If H$_2$ and HI column densities in the inner disk become optically thin in transitional disks, more stellar Ly$\alpha$ can irradiate molecular material in the outer disk and may explain the observed correlation between r$_{out}$(H$_2$) and n$_{13-31}$. This suggests that HI clearing from the inner disk may happen over a similar timescale as the characteristic dust dissipation \citep{Wyatt+08, Ribas+14} and mass accretion quenching \citep{Fedele+10}. This inner-to-outer disk dissipation is in agreement with the UV switch model, which describes the dispersal of inner disk gas cut off from the gas reservoir of the outer disk, due to selective photoevaporation of material out to r $\sim$ 5 - 10 AU \citep{Clarke+01, Alexander+06}. Observations of other outer-disk molecules photo-excited by Ly$\alpha$ radiation provide additional evidence for the loss of HI in the inner disks of transitional objects. For example, \citet{Schindhelm+12a} observe FUV-CO fluorescence, also powered by stellar Ly$\alpha$-pumping, at T$_{exc}$ $\sim$ 500K, in transitional phase objects with an average emission radius R$_{CO}$ $\sim$ 1 - 5 AU. This indicates that less HI and H$_2$ column is present in the inner disk to shield the stellar Ly$\alpha$ flux from reaching the cooler CO material at intermediate radii in transition systems. \\
\indent Figure \ref{fig:radial_comparisons} shows a 1D radial comparison of dust and molecular tracers determined in our targets. We present the locations of the outer radiation boundary for H$_2$ FUV emission, as determined from our models (r$_{out,[1,4]}$; green triangles), and the observed dust cavity walls of the transitional disk populations (r$_{cavity}$; blue squares). For all transitional disks, we find r$_{out,[1,4]}$ inward of r$_{cavity}$, meaning that the H$_2$ population observed in all transition PPDs radiates where the dust is optically thin, suggesting that the H$_2$ populations remain optically thick even after the dust grains have dissipated. Studies like \citet{vanderMarel+15} also find a substantial depletion of the dust-to-gas ratio inside the dust cavities of well-studied transition disks, which is consistent with our findings. 
%
%
%
%
%
%
\subsection{Near-IR CO Emission and Comparison to Snow Line Radii}
\begin{figure*}[htp]
	\includegraphics[angle=90, width=1.0\textwidth]{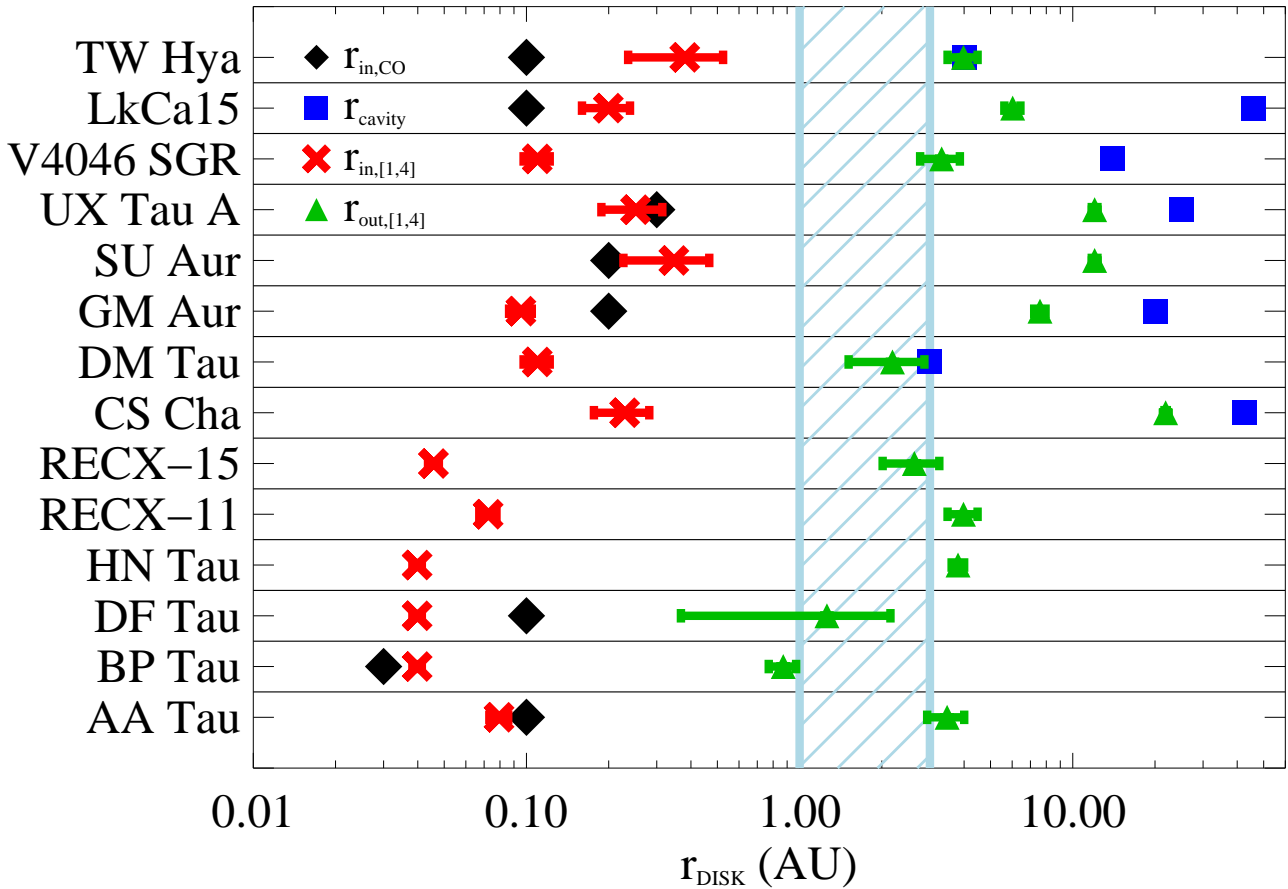}
	\caption{A radial comparison of the inner and outer extent of FUV H$_2$ emission (this work), the innermost radius of near-IR CO emission \citep{Salyk+11b}, and dust cavity locations in transition disk targets (see Table \ref{tab:disk_parms} for references). The light blue shaded area from 1 AU $\leq$ r$_{DISK}$ $\leq$ 3 AU represents the theoretical water-ice snow line for the presence of water-ice at the midplane of primordial and transitional PPDs \citep{Baillie+15}.}
	\label{fig:radial_comparisons}
\end{figure*}
%
%
%
Figure \ref{fig:radial_comparisons} includes radial estimates of the inner radiation boundary for H$_2$ FUV emission (r$_{in,[1,4]}$; blue x's) and the inner radius of near-IR CO emission, determined from LTE models presented by \citet{Salyk+11b} (r$_{in,CO}$; black diamonds). The inner disk emission radii of FUV H$_2$ and near-IR CO appear to be roughly co-spatial, which is a result also found by \citet{France+12b} when comparing the observed FWHMs of FUV H$_2$ fluorescence emission and near-IR CO emission. An extensive study by \citet{Brown+13} concluded there is a correlation between the near-IR CO P(8) equivalent width and dust disk dispersal in transitional disks, suggestive of outer radial origins of the CO emission as PPD dust evolves. We have shown that r$_{in,[1,4]}$ increases with n$_{13-31}$ and decreases with $\dot{M}$, providing further evidence that the inner gas disk environment becomes optically thin as disks evolve towards the transition stage. \\
\indent We note the disk locations of possible theoretical snow lines in PPDs and these radii coincide with the H$_2$ fluorescence in Figure \ref{fig:radial_comparisons}. As the disk evolves, it cools over time, so the snow line is expected to migrate inward in the disk as the protostellar system ages \citep{Cassen+94}. Several independent studies (e.g. \citealt{Meijerink+09, Mandell+12}) conclude that the location of the water-ice snow line in PPDs are expected to be found within r $\sim$ 1 - 3 AU for all PPD states. \citet{Baillie+15} shows that the evolution of the water-ice snow line at all stages of PPD evolution (from ages 10$^6$ - 10$^7$ yr) only varies by $\sim$ 0.5 AU. Observations of H$_2$O and OH (which is thought to be a bi-product of H$_2$O photo-dissociation) in the near- and mid-IR are also consistent with these condensation radii \citep{Malfait+98, Carr+04, Mandell+08, Salyk+08}. Figure \ref{fig:radial_comparisons} includes a shaded blue region that represents the assumed generalized H$_2$O snow line radii in PPDs, located between r$_{DISK}$ = 1 - 3 AU. With the exception of BP Tau and DF Tau A, all targets have outer H$_2$ emission radii that extend to within or outward of the water-ice snow line. \\
%

\section{Conclusions and Future Work}

We have created 2D radiative transfer models of FUV H$_2$ fluorescence emission in PPDs and compared them with observations made with \textit{HST}-COS and STIS. We analyze the radial distribution of H$_2$ emission produced by parametrized models, which are determined using a reduced-$\chi ^2$ statistic, to understand how the emitting H$_2$ regions changes as PPD dust disks evolve. We summarize our findings and interpret the evolutionary behavior of the molecular disk atmosphere as the inner dust disks of PPDs disperse:
\begin{enumerate}
	\item The modeled H$_2$ radial distributions differ between primordial and transitional disks. Primordial disks have the majority of the total H$_2$ flux arising from the innermost disk radii and less produced outside $\sim$ 1 AU. For transitional disks, the total H$_2$ flux migrates to larger disk radii, producing less flux in the innermost disk and more out to r $\sim$ 10 AU.
	\item We see a positive correlation between the resulting inner and outer emission radii of FUV H$_2$ (r$_{in}$ and r$_{out}$), which supports the result described in conclusion 1. This can mean: a) that the physical structure (i.e., temperature) of the warm molecular disk atmosphere changes as PPDs evolve, b) the warm, ground-state H$_2$ populations [$v$,$J$] change, resulting in evolving regions of the disks where the warm H$_2$ atmosphere will reprocess the stellar Ly$\alpha$ radiation field, or c) H$_2$ is being destroyed in the inner disk and not re-formed, owing to the lack of dust grains; the latter point allows stellar Ly$\alpha$ to penetrate to larger r$_{out}$.
	\item We observe positive correlations between r$_{in}$, r$_{out}$, and n$_{13-31}$, suggesting that r$_{in}$ corresponds with the loss of warm, small dust grains in the innermost disk. We find a negative correlation between r$_{in}$ and $\dot{M}$, providing evidence that the warm H$_2$ inner disk atmosphere may be physically thinned or cleared as the PPDs evolve, possibly by the loss of a molecular formation site as the dust grains dissipate from the atmosphere. Using the observed dust cavity radii of the transitional disk targets, we compare r$_{out}$ to r$_{cavity}$ and find that, for all transition disk targets, r$_{out}$ is found inward of r$_{cavity}$. This indicates that the warm H$_2$ disk (for r $>$ r$_{in}$) remains optically-thick where the warm dust grains are optically-thin in the disks. This suggests that the physical mechanism that clears or settles the inner disk dust either does not have the same effect on the molecular disk atmosphere, or there is a time lag for the gas disk to respond to the changes observed in the dust distribution.
	%
	%
	\item We examine where the emitting H$_2$ originates in the disk relative to warm CO and the theoretical location of water-ice snow lines. Inner disk CO is roughly co-spatial with r$_{in}$ for all targets, which could point to the dispersal of the warm molecular disk atmospheres of evolving disk systems. With the exception of a few primordial disk targets, all targets have emitting H$_2$ regions that encapsulate the theoretical water-ice snow line. If disk clearing mechanisms, such as disk photoevaporation via EUV/X-ray photons, are primarily responsible for the final dispersal of the gas disk at the end of the PPD lifetime, it is important to examine late-type PPDs to monitor molecular disk clearing as transitional disks evolve to debris disks.
\end{enumerate}
%
\acknowledgments
This research was funded by the NASA Astrophysics Research and Analysis (APRA) grant NNX13AF55G, \textit{HST} GO program 12876, and \textit{HST} AR program 13267, and uses archival NASA/ESA \textit{Hubble Space Telescope} observations, obtained through the Barbara A. Mikulski Archive for Space Telescopes at the Space Telescope Science Institute. RDA acknowledges support from The Leverhulme Trust through a Philip Leverhulme Prize, and from the UK Science \& Technology Facilities Council (STFC) through Consolidated Grant ST/K001000/1. PCS gratefully acknowledges support from the ESA Research Fellowship. KH thanks Katherine Rosenfeld for early discussion about the disk modeling geometry; Eliot Kersgaard for help with early analysis and line fitting tools; and Susan Edwards, Hans Moritz Guenther, and Laura P$\acute{e}$rez for productive discussions about the H$_2$ radiation field results at the $225^{th}$ AAS conference in Seattle, WA. We thank the anonymous referee for helpful comments.



\end{document}